\documentclass[twocolumn]{aastex631}
\usepackage{amsmath,graphicx,hyperref}
\usepackage{natbib,threeparttable,rotating}
\usepackage{ulem}
\usepackage{multirow}
\newcommand{\etal}{et al.}

\begin{document}

\title{Fast and Flexible Inference Framework for Continuum Reverberation Mapping using Simulation-based Inference with Deep Learning} 

\author[0000-0002-0311-2812]{Jennifer~I-Hsiu Li}
\affiliation{Michigan Institute for Data Science, University of Michigan, Ann Arbor, MI, 48109, USA}
\affiliation{Department of Astronomy, University of Michigan, Ann Arbor, MI, 48109, USA}

\author[0000-0001-9487-8583]{Sean~D.~Johnson}
\affiliation{Department of Astronomy, University of Michigan, Ann Arbor, MI, 48109, USA}

\author[0000-0001-8868-0810]{Camille Avestruz}
\affiliation{Department of Physics, University of Michigan, Ann Arbor, MI, 48109, USA}
\affiliation{Leinweber Center for Theoretical Physics, University of Michigan, Ann Arbor, MI, 48109, USA}

\author[0000-0002-5386-7076]{Sreevani Jarugula}
\affiliation{Fermi National Accelerator Laboratory, Batavia, IL 60510, USA}

\author[0000-0003-1659-7035]{Yue Shen}
\affiliation{Department of Astronomy, University of Illinois at Urbana-Champaign, Urbana, IL 61801, USA}
\affiliation{National Center for Supercomputing Applications, University of Illinois at Urbana-Champaign, Urbana, IL 61801, USA}

\author[0000-0001-6846-9399]{Elise Kesler}
\affiliation{Department of Astronomy, University of Michigan, Ann Arbor, MI, 48109, USA}

\author[0000-0002-2662-9363]{Zhuoqi (Will) Liu}
\affiliation{Department of Astronomy, University of Michigan, Ann Arbor, MI, 48109, USA}

\author[0000-0002-9141-9792]{Nishant Mishra}
\affiliation{Department of Astronomy, University of Michigan, Ann Arbor, MI, 48109, USA}

\shorttitle{CRM with SBI}
\shortauthors{Li \etal}

\begin{abstract} 
{Continuum reverberation mapping (CRM) of active galactic nuclei (AGN) monitors multiwavelength variability signatures to constrain accretion disk structure and supermassive black hole (SMBH) properties.} The upcoming Vera Rubin Observatory's Legacy Survey of Space and Time (LSST) will survey tens of millions of AGN over the next decade, with thousands of AGN monitored with almost daily cadence in the deep drilling fields. However, existing CRM methodologies often require long computation time and are not designed to handle such large amount of data. In this paper, we present a fast and flexible inference framework for CRM using simulation-based inference (SBI) with deep learning to estimate SMBH properties from AGN light curves. We use a long-short-term-memory (LSTM) summary network to reduce the high-dimensionality of the light curve data, and then use a neural density estimator to {estimate the posterior of SMBH parameters}. Using simulated light curves, we find SBI can produce more accurate SMBH parameter estimation with $10^3-10^5$ times speed up in inference efficiency compared to traditional methods. {The SBI framework is particularly suitable for wide-field RM surveys as the light curves will have identical observing patterns, which can be incorporated into the SBI simulation.} We explore the performance of our SBI model on light curves with irregular-sampled, realistic observing cadence and alternative variability characteristics to demonstrate the flexibility and limitation of the SBI framework. 
\end{abstract}



\section{Introduction}

\begin{figure*}
    \centering
    \includegraphics[trim=0 110 80 0, clip, width=\textwidth]{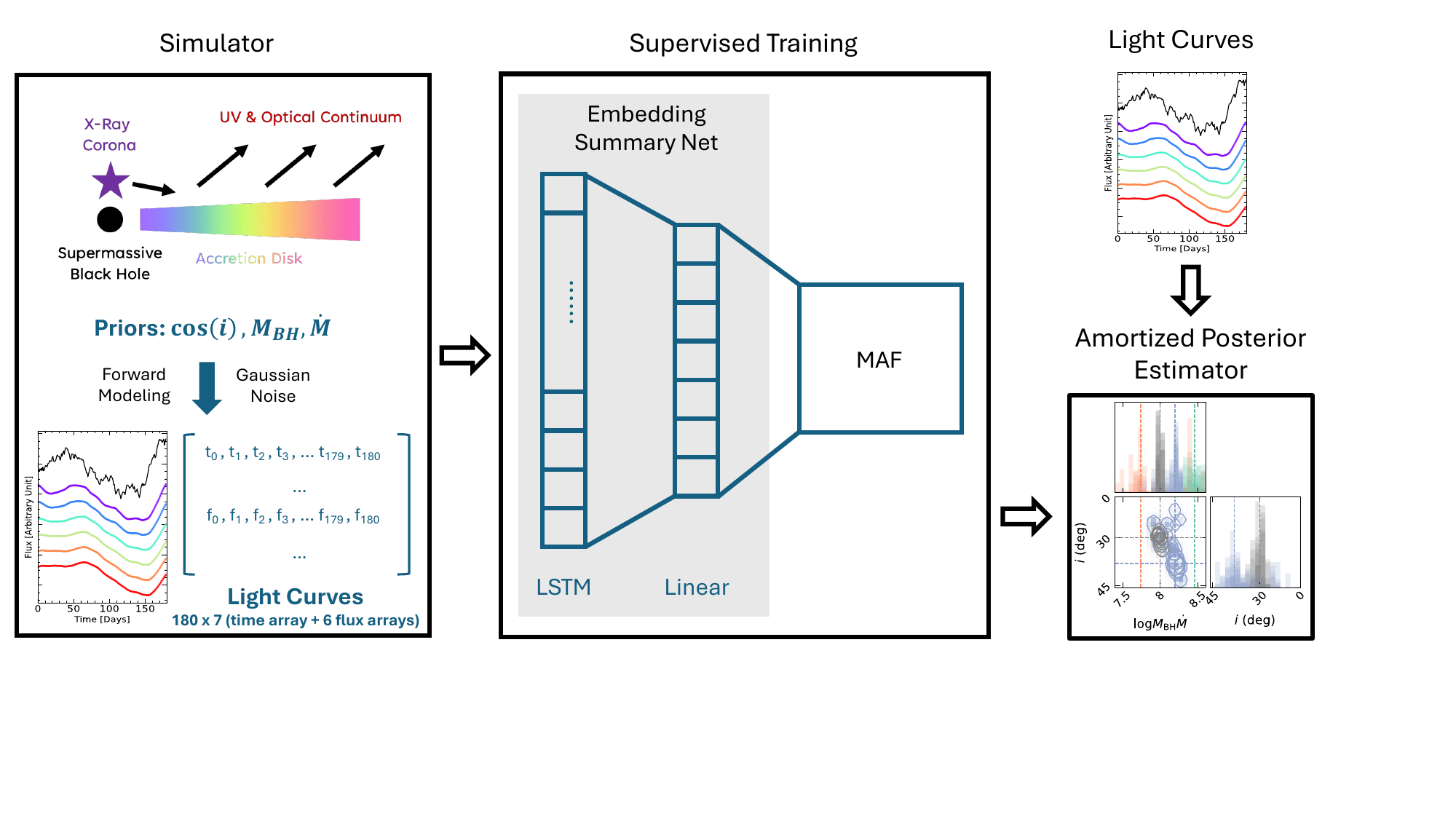}
    \caption{An overview of the SBI setup of this work. The simulator includes the accretion disk model and light curve generation described in Section \ref{sec:sim}. {We train a shallow embedding summary network and masked autoregressive flow (MAF) density estimator to evaluate the BH parameters (cosine of the accretion disk inclination cos($i$), BH mass $M_{\rm BH}$, and  accretion rate $\dot{M}$) using the 180$\times$7 light curve arrays (1 array for time $t$ and 6 arrays for the multiwavelength flux $f$, each with 180 time steps) (Section \ref{sec:model}). We implement the SBI framework on idealized (Section \ref{sec:overall_result}) and realistic (Section \ref{sec:discuss_gap} and \ref{sec:discuss_drw}) simulated light curves.}}
    \label{fig:overview}
\end{figure*} 

While it is extremely difficult to spatially resolve the innermost regions of an active galactic nuclei (AGN), its geometry, dynamical structure, and photoionization properties are encoded in AGN variability \citep{VandenBerk_etal_2004, MacLeod_etal_2010}. In the simple ``lamp post'' model, as X-ray/UV emission from the AGN center propagates outward, it is absorbed, reprocessed, and re-emitted in different parts of the AGN spectra, corresponding to distinct structures of the AGN (e.g., accretion disk, broad and narrow line regions, dusty torus, etc.). The variability signatures are also propagated, but delayed by the light-traveling time.  By monitoring the delayed response between optical photometric bands, we can map out the temperature and size profiles of an accretion disk, which is known as the (continuum) reverberation mapping technique \citep[CRM, for a recent review, see][]{Cackett_etal_2021}. CRM can provide measurements of the accretion disk structure and BH properties, which are difficult to measure even for the most nearby SMBHs.

However, CRM has only been applied to a few hundreds of objects to date \cite[e.g.,][]{Sergeev_etal_2005, DeRosa_etal_2015, Jiang_etal_2017, Homayouni_etal_2019, Edelson_etal_2019, Yu_etal_2020}, due to stringent observing requirements. Typical accretion disk sizes are on the scales of a few light-days, so a successful monitoring CRM campaign usually requires weeks of almost daily observations to ensure distinctive variability trends are captured.

The upcoming Vera Rubin's Observatory Legacy Survey of Space and Time \citep[LSST,][]{Ivezic_etal_2019} will survey the night sky for the next decade in six photometric bands, monitoring tens of million AGNs on days-to-weeks timescale. Roughly 6\% of LSST's total survey time will be dedicated to providing deeper and higher cadence monitoring in five deep drilling fields (DDFs). With proposed almost daily cadence \citep{Brandt_etal_2018} and $\sim$3000 X-ray and SED-selected AGN or AGN candidates per field \citep{Ni_etal_2021a, Zou_etal_2022}, LSST's DDF will revolutionize accretion disk and AGN research by providing an unprecedented amount of high quality light curves.  

Another major bottleneck for CRM surveys is the long computation time associated with traditional lag-measuring software, especially fitting codes relying on Markov Chain Monte Carlo (MCMC) methods like {\tt JAVELIN} \citep{Zu_etal_2011} and {\tt CREAM} \citep{Starkey_etal_2016}. Due to the large number of parameters to model AGN light curves, these MCMC codes typically take up to hours to converge for a single target. These traditional methods will not feasibly deal with the vast amount of data from future surveys like the LSST. Many recent works have turned to machine learning and deep learning techniques to provide more efficient algorithms for AGN science, e.g., modeling AGN variability \citep{Tachibana_etal_2020}, inferring BH parameters \citep{Fagin_etal_2024}, and detecting anomalous light curves \citep{Sanchez-saez_etal_2021}.

In this paper, we propose the use of simulation-based inference \citep[SBI,][]{Cranmer_etal_2020} with deep learning to infer SMBH properties in the era of LSST. The main advantage of SBI is inference speed; once the SBI model is trained, inference on new data can be drawn with negligible computation time without retraining the model {(i.e., amortized posterior)}. In addition, the SBI model can be trained on different accretion disk and BH models, making it a flexible framework to compare physical assumptions. Finally, SBI is particularly efficient for large, uniform surveys like the LSST DDF, since all light curves from each DDF will have near identical observing patterns (e.g., cadence, weather gap, flux uncertainties) that can be directly integrated into the model training. 

As a proof-of-concept, we train our SBI model on idealized simulated light curves, and then test its inference accuracy and computation speed against other traditional methods in this paper. Figure \ref{fig:overview} shows the overview of this framework.  We describe our simulations in Section \ref{sec:sim} and our inference methods in Section \ref{sec:method}. The main results are presented in Section \ref{sec:result}, including the overall performance of SBI and its comparison with traditional methods. We discuss more realistic and non-ideal test cases in Section \ref{sec:discussion} and conclude in Section \ref{sec:conclusion}.

\begin{figure}
    \centering
\includegraphics[width=0.48\textwidth]{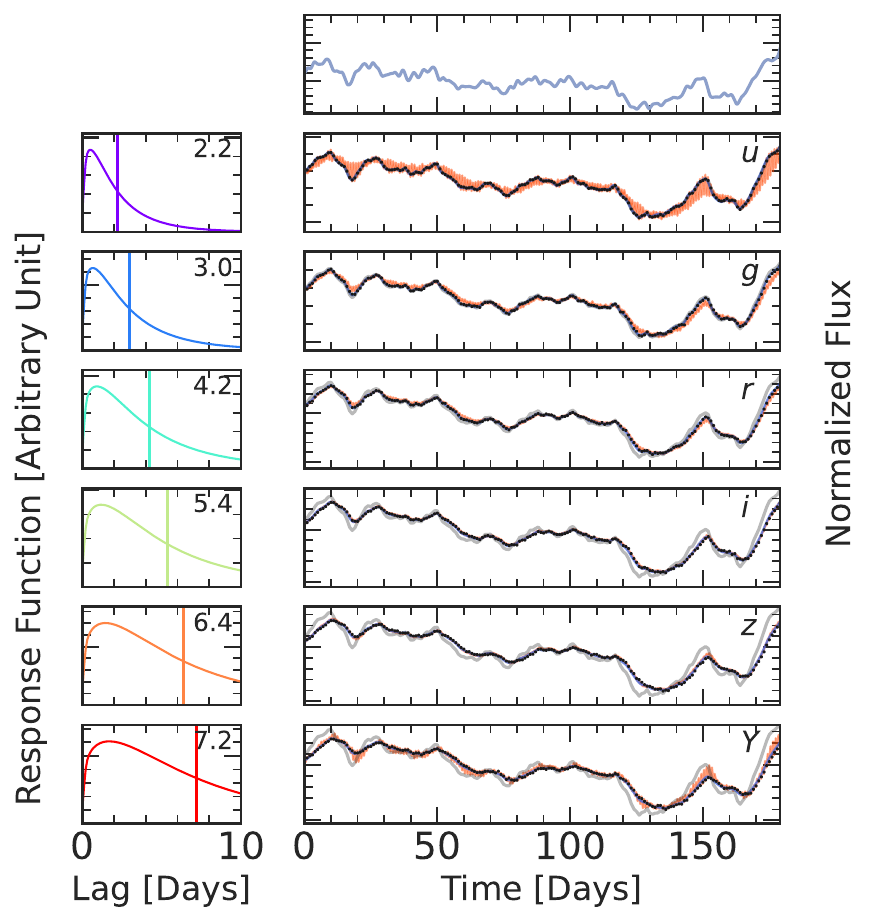}
    \caption{{{\it Left} panels: The response functions (Equation \ref{eq:response}) in the 6 LSST bands of an AGN with $M_{\rm BH}=10^8 {\rm M}_{\rm \odot}$, $\dot{M}=1\,{\rm M}_{\rm \odot}{\rm yr}^{-1}$ and $i=30^{\circ}$. The centroid time delays of each response function are shown in the vertical lines and labeled at the top right corner of each panel. {\it Right} panels: An example of the simulated light curves (black dots, from the second to the last panels). The shaded light curves are the best fit models (see Section \ref{sec:compare}) from {\tt JAVELIN} (orange) and {\tt CREAM} (blue), and the top panel shows the best-fitted driving light curve by {\tt CREAM}. The grey line in each panel shows the $u$ band light curve as reference. The longer wavelength light curves are the convolution of the $u$ band light curve and the response functions.}}
    \label{fig:lc}
\end{figure}

\section{Simulations}\label{sec:sim}
\subsection{Accretion Disk Model}
The accretion disk is modeled as a standard optically thick, geometrically thin disk from \cite{Shakura_Sunyave_1973}, following the simulation setup used in \cite{Starkey_etal_2016, Starkey_etal_2017}. The accretion disk temperature profile $T(r)$ is determined by two heating mechanisms: viscous heating from the differential disk rotation and irradiation from the X-ray corona located above the center SMBH   \citep{Frank_etal_2002}, and it is described by:
\begin{equation}\label{eq:temp}
    {T(r)}^{4} = \frac{3GM_{\rm BH}\dot{M}}{8\pi\sigma r^{3}}\left(1-\sqrt{\frac{r_{\rm in}}{r}}\right)+
    \frac{L(1-a)h_x}{4\pi\sigma (r^2+{h_x}^2)^{3/2}}
\end{equation}
where $G$ is the gravitational constant, ${M}_{\rm BH}$ is the black hole mass, $\dot{M}$ is the accretion rate, $\sigma$ is the Stefan–Boltzmann constant,  $L$ is the bolometric luminosity and $L=\eta \dot{M} c^2$ with typical radiative efficiency $\eta=0.1$, $a$ is the disk albedo, $h_x$ is the X-ray corona height, and $r_{\rm in}$ is the inner most stable circular orbit (ISCO) of the black hole. Here, we assume $r_{\rm in} = 3 r_s$ for a non-rotating SMBH, where $r_s$ is the Schwarzschild radius, $h_x=6r_s$ for a X-ray corona relatively close to the center SMBH, and $a=0$. The radiation at a certain wavelength $\lambda$ from a given radius $r$ on the accretion disk can then be described using the Planck function: 
\begin{equation}\label{eq:planck}
    B_{\lambda}(\lambda,T) = \frac{2hc}{\lambda^5}\frac{1}{e^{hc/\lambda kT}-1}
\end{equation}
where $h$, $k$, $c$ are the Planck constant, Boltzmann constant, and the speed of light, respectively.

The total flux from the accretion disk at a certain time $t$ is the integral of the Planck function over the disk surface: 
\begin{equation}\label{eq:tot_flux}
    F_\lambda(\lambda, t) = \int B_{\lambda}(\lambda,T(t-\tau)) d\Omega
\end{equation}
where $T(t-\tau)$ is the disk temperature at the look-back time $\tau$, i.e., the light traveling time between the variable irradiation source to the accretion disk at radius of $r$:
\begin{equation}
    \tau = \frac{r}{c}(1-\cos \theta \sin i)
\end{equation}
for a flat disk. $\theta$ is the azimuth angle of the reprocessing location on the accretion disk, and $i$ is the inclination angle of the disk with respect to the observer ($i=0$ for a face-on disk).  
 
Assuming $L(t)$ is the driving light curve, the integral in Equation \ref{eq:tot_flux} can be rewritten as 
\begin{equation}
    F_\lambda(\lambda, t) = \bar{F}(\lambda)+\Delta F(\lambda)\int_{0}^{t} \psi(\tau|\lambda)L(t-\tau) d\tau
\end{equation}
Here, $F$ is separated into a background component $\bar{F}(\lambda)$ and a variable component $\Delta F(\lambda)$, and $\psi$ is the disk response function
\begin{equation}\label{eq:response}
    \psi(\tau|\lambda) = \int d\Omega \frac{\partial B_{\lambda}(\lambda,T)}{\partial T} \frac{\partial T}{\partial L} \frac{\partial L}{\partial F} \delta(\tau-\tau') 
\end{equation}
where $\Delta \Omega = r \Delta r \Delta \theta / D_L^2$, with luminosity distance $D_L$. The differentials $\partial B/\partial T$ and $\partial T/\partial L$ can be derived using Equations \ref{eq:planck} and \ref{eq:temp}. $\partial L/\partial F$ introduces a constant $4\pi$ to the equation, and the delta function ensures only the narrow annulus at $\tau$ is contributing to the disk reprocessing. In practice, we use a narrow Gaussian distribution with a width of 0.1 days to approximate the delta function and ensure smoothness of the response function. We evaluate the response function up to $\tau=30$ days and normalize the response function so that $\sum\psi(\tau|\lambda)\delta\tau=1$. The disk response function is a function of $M_{\rm BH}\dot{M}$ and $i$. Figure \ref{fig:lc} (left panels) shows how the response function and centroid time lag change at different wavelengths for a typical SMBH of $M_{\rm BH}=10^8 {\rm M_\Sun}$ and $\dot{M} = 1 {\rm M_\Sun yr^{-1}}$ at $i=30^{\circ}$.

\subsection{Simulated Light Curves}\label{sec:sim_lc}
The AGN light curves on day-to-week timescales in the optical wavelength can be modeled by the damped random walk \citep[DRW,][]{Kelly_etal_2009, Kelly_etal_2011, Kozlowski_etal_2010}. DRW light curves are described by two parameters, the rms variability on long timescale, ${SF}_{\rm inf}$, and the damping timescale, ${\tau}_{\rm DRW}$, at which the light curve becomes uncorrelated. To generate simulated observations, we first generate a sample of 10,000 mock BH with uniform distribution of $\log{M}_{\rm BH}$, $\log{\dot{M}}$, and cos($i$) (see Table \ref{tab:parameters} for the prior range), and then use the empirical relations from \cite{MacLeod_etal_2010} to generate the DRW parameters for a driving light curve. Since the \cite{MacLeod_etal_2010} relations are only fitted from the SDSS optical bands, we assume the DRW parameters for the driving light curve is similar to the shortest observed wavelength at $u$ band. 

Next, we convolve the driving light curve with the transfer function (Equation \ref{eq:response}) to generate $u, g, r, i, z, Y$ light curves for 180 days on a 1-day cadence with no gaps {as a simplified benchmark example. We will explore more realistic cadences in Section \ref{sec:discuss_gap}.} We perturb each data point with the flux level as mean, and 1\% flux level as standard deviation to simulate 1\% measurement uncertainty. {This is also a simplified assumption, the actual measurement uncertainty will depend on the brightness of the targets and observing strategy of the survey.} Figure \ref{fig:lc} (right panels) shows an example of the driving light curve and the convolved multiband light curves. For each set of BH parameters, we generate 10 light curve realizations using the same DRW parameters to improve the variety of possible observations for each set of BH parameters (also known as the data augmentation process), creating a training set of 100,000. For evaluating the performance of our models, we generate an independent testing data with 1,000 unique set of black holes parameters with 1 light curve realization each and 1\% measurement uncertainty.

\begin{table*}
\hspace{-2cm}
    \centering
    \begin{tabular}{lrrrr}
    \hline\hline
      Parameter   & Prior Range (Step) & Prior Distribution & Default &Best \\
    \hline
    {\it Simulation Parameters} &&&&\\
      cos($i$) & [cos(0), cos($\pi$/4)] & Uniform & -- & -- \\
      ${\rm M}_{\rm BH}$ & [$10^{5}$,$10^{10}$] & Log & -- & -- \\
      ${\rm \dot{M}}$ & [$10^{-2}$,$10^{2}$] & Log & -- & --\\
    \hline
    {\it Neural Network Parameters} &&&&\\
      LSTM units  & [16,128] (16) & Uniform & 64 & 128 \\
      LSTM out & [16,128] (16) & Uniform & 16 & 48 \\
      MAF hidden features  & [8,64] (8) & Uniform & 25 & 48\\
      MAF transforms & [8,64] (8) & Uniform & 10 & 16\\
      Learning rate  & [$10^{-3}$,$10^{-5}$] & Log & $0.005$ & 0.001 \\
      Batch size  & [16,128] (8) & Uniform & 50 & 80 \\
    \hline
    & & $\sigma$(cos($i$))  & 0.01 & 0.01 \\
    & & $\sigma$($\log{M}_{\rm BH}$) & 0.08 & 0.08 \\
    & & $\sigma$($\log{\dot{M}}$) & 0.08 & 0.08 \\
    \hline\hline
    \end{tabular}
    \caption{Simulation and SBI model parameter range and priors. {The six hyperparameters optimized with {\tt optuna} are the numbers of the LSTM units, output features from the linear layer, hidden features and layers in MAF, and the training batch size and learning rate. } {The last three rows show the 1$\sigma$ uncertainty of each BH parameters.}}
    \label{tab:parameters}
\end{table*}   

\section{Inference Methods}\label{sec:method}
\subsection{Simulation-based inference}\label{sec:model}
SBI \citep{Cranmer_etal_2020} are inference methods that compute the posterior by comparing an observation to simulations, often to solve problems with intractable likelihoods. In the context of machine learning, SBI utilizes deep neural networks to ``learn'' the posterior distribution from input physical priors and simulated data generated from a simulator. In this work, the input physical parameters are BH mass $M_{\rm BH}$, accretion rate $\dot{M}$, and cosine of the accretion disk inclination $cos(i)$, and the simulated observation is the daily-sampled multi-band light curves generated as described in Section \ref{sec:sim_lc}. For the inference algorithm, we use the Sequential Neural Posterior Estimation \citep[SNPE, or Automatic Posterior Transformation, APT,][]{SNPE} implemented in the Python SBI toolkit {\tt sbi} \citep{sbi}.

Due to the high dimensionality of the light curve data, we first use a long-short-term memory \cite[LSTM, ][]{LSTM} embedding network to extract the light curve features. LSTM is a recurrent neural network (RNN) architecture that can retain time-dependent properties of the data by modeling each data point sequentially using the current time step (short-term memory) and previous time steps (long-term memory). Since AGN light curves can be loosely modeled by autoregressive models, the LSTM network is more efficient in learning their characteristics than a non-sequential neural network (e.g., CNN, see Appendix \ref{app:other_method} for comparison). We opt to use {a shallow LSTM network for the embedding summary net, including one LSTM layer and one linear layer} to map the LSTM outputs to the chosen number of output features. These features are then feed into a neural density estimator to learn the {correlation between the input BH parameters and the corresponding simulation.} We use the Masked Autoregressive Flow \citep[MAF,][]{Papamakarios_etal_2017} in {\tt sbi}, which is a type of normalizing flow designed to transform simple distributions to complex distributions. {An approximate posterior over the entire trained prior range is built based on the trained neural network, and the posterior for a particular observation (i.e., new light curves) can be drawn without retraining the neural network, which is known as amortized posterior.} Figure \ref{fig:overview} (middle portion) illustrates the neural network setup in this paper, and {the detailed architecture is listed in Appendix \ref{app:other_method}}. 

To determine the best configurations of the neural network, we use {\tt optuna} \citep{optuna} to optimize six hyper-parameters: the width of the LSTM network, the number of output features from the linear layer, the numbers of hidden features and layers in MAF, and the training batch size and learning rate. {\tt optuna} uses Bayesian optimization to efficiently search for the optimal hyperparameter setup within the search range by maximizing the {validation probability}. Due to the long training time from SBI, we use only 30,000 light curve sets from the training set and search over a coarse grid of the parameters to save time. The tested parameter range, prior, and the best hyperparameters are listed in Table \ref{tab:parameters}.
{Over the completed 30 {\tt optuna} trials, we find that the number of LSTM units is the most important hyperparameter, while the rest of the hyperparameters do not make a big difference in validation probability, which is similar for all {\tt optuna} runs with more than $\sim$50 LSTM units.} {The inference accuracy from the best {\tt optuna} model are similar to our default SBI model. For the rest of the paper (including Sections \ref{sec:overall_result}, \ref{sec:discuss_gap}, \ref{sec:discuss_drw}), we show results using the default SBI parameters listed in Table \ref{tab:parameters}.}

\begin{figure*}
    \centering
    \includegraphics[width=\textwidth]{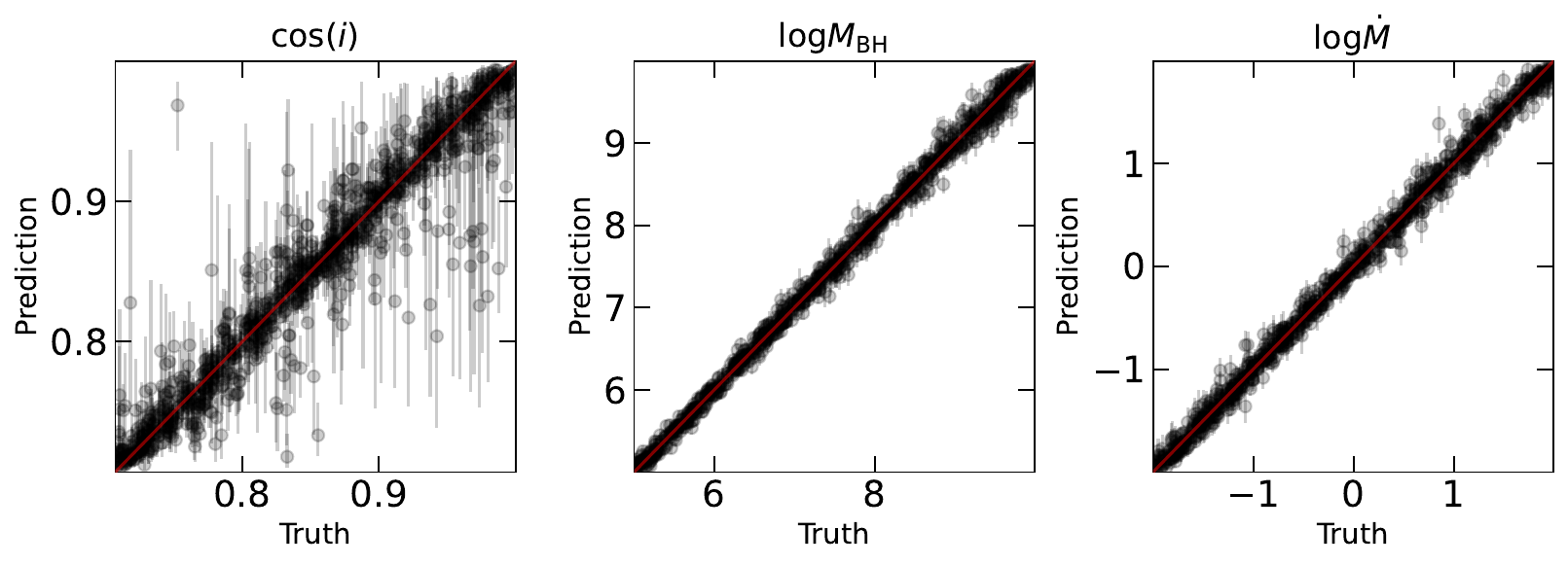}
    \includegraphics[width=\textwidth]{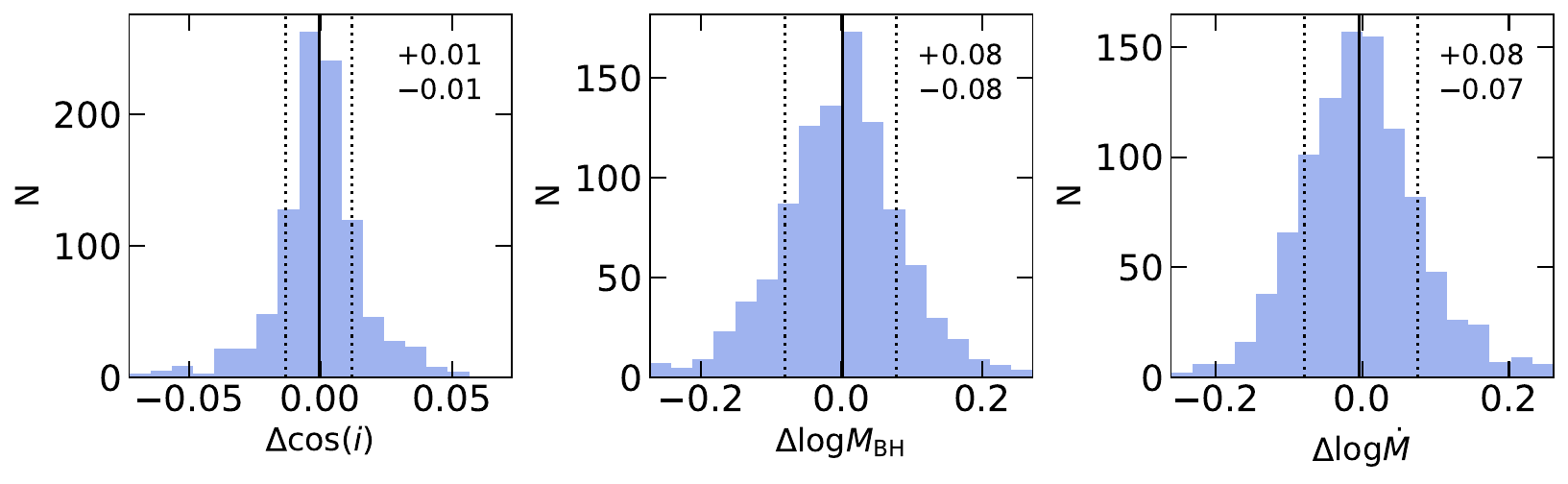}
    \includegraphics[width=\textwidth]{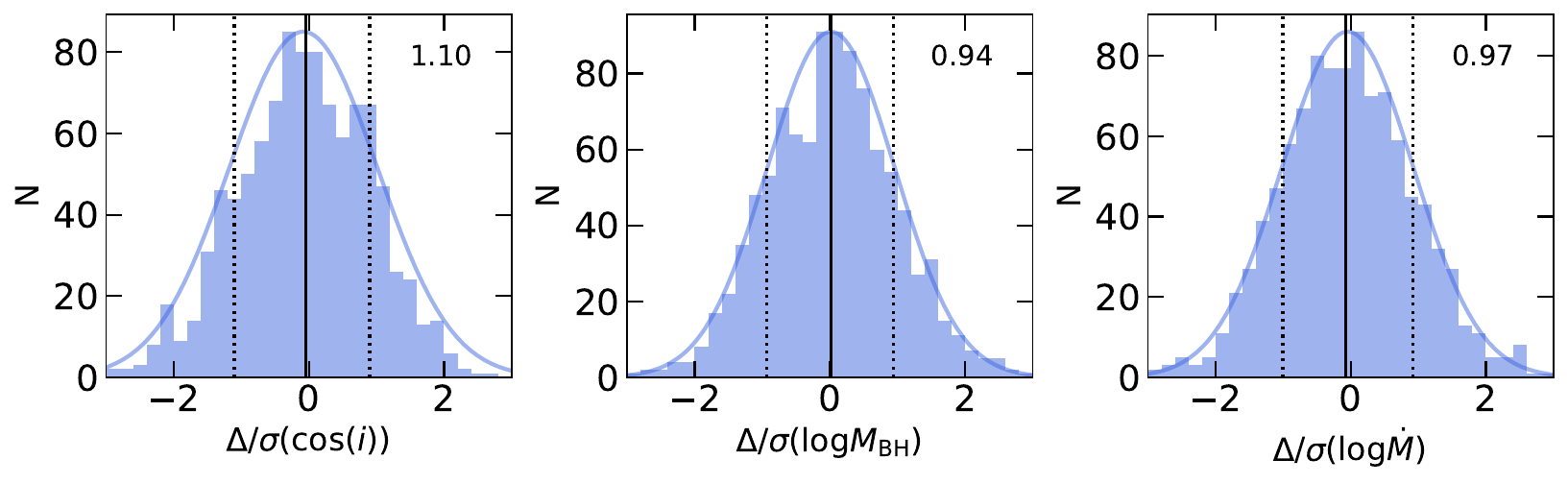}
    \caption{Evaluation results of testing data. The {\it top} row shows the comparison between the true and predicted values, the {\it middle} row shows the distribution of the difference between the true and predicted values, and the {\it bottom} row shows the distribution of the difference between the true and predicted values, normalized by the {$1\sigma$ uncertainties of the SBI posterior}. The red lines in the top row trace the 1:1 relation, the solid (dotted) black lines in the middle and bottom rows denote the median (and 16, 84 percentile) of the distributions. The blue curves in the bottom panels show the best-fit Gaussian, with the width of the Gaussian noted in the upper right corner. The trained SBI model can constrain $M_{\rm BH}$ and $\dot{M}$ to $\sim$0.08\,dex, and $cos(i)$ to 0.01 (a few degrees in the prior range $i=0-45^\circ$).}
    \label{fig:sbi_result}
\end{figure*}

\subsection{Traditional Inference Methods}
To assess the performance of SBI, we compare our results with three commonly-used RM methods, the interpolated cross-correlation function (ICCF), {\tt JAVELIN}, and {\tt CREAM}. We briefly describe each method and their respective parameter setups in this section. 

\subsubsection{ICCF}
ICCF \citep{Peterson_etal_1998} is the most commonly-used method of measuring lags for RM. ICCF measures lags by searching for the peak cross-correlation of two light curves shifted over a grid of time-lags. We implement the ICCF measurements using the code {\tt PyCCF} \citep{PyCCF}. We measure the lag using the centroid of the CCF peak, calculated over a search range of $\pm30$\,days with time steps of 0.2\,days. We perform 500 Monte Carlo (MC) simulations with the traditional flux randomization and random subset
sampling (FR/RSS) procedure, and adopt the 50 (16, 84) percentiles as the measured lag (and its 1$\sigma$ uncertainties).

To convert from lag to accretion disk size and BH parameters, we define the flux-weighted accretion disk size $R_\lambda$ at wavelength $\lambda = Xhc/kT$, and equation \ref{eq:temp} becomes 
\begin{equation}\label{eq:disksize}
    T^4 = (\frac{Xhc}{\lambda k})^4 = \frac{3GM_{\rm BH}\dot{M}(1+\kappa)}{8 \pi \sigma R_{\lambda}^3},
\end{equation}
where we assume $R_\lambda>>r_{in}$. $X$ is the correction factor that accounts for the different disk annuli that contributed to the emission at $\lambda$. We adopt $X=2.49$ following \cite{Fausnaugh_etal_2016}. $\kappa$ is the ratio between external and local heating (the two terms in equation \ref{eq:temp}). {Assuming negligible external heating (i.e., $\kappa=0$)}, disk size is a power-law of wavelength
\begin{equation}\label{eq:lag_disksize}
   c\tau = R_{\lambda} = R_0 (\frac{\lambda}{\lambda_0})^\beta ,
\end{equation}
{where $R_0$ is the disk size at the reference wavelength $\lambda_0$, chosen to be 4000\AA\,here}. We fix $\beta = 4/3$ for a thin disk model, and fit the ICCF lags at each wavelengths to calculate ${M}_{\rm BH}{\dot{M}}$, as ${M}_{\rm BH}$ and ${\dot{M}}$ are completely degenerate in Equation \ref{eq:disksize}.

\subsubsection{{\tt JAVELIN}}
The code {\tt JAVELIN}\footnote{https://github.com/nye17/javelin}\footnote{https://github.com/legolason/javelin-1} \citep{Zu_etal_2011, Zu_etal_2016} assumes the light curves at longer wavelengths are shifted, stretched, and scaled versions of a shorter wavelength light curve (the driving light curve). {\tt JAVELIN} employs a two-step MCMC fitting. It first fits the shortest wavelength light curve with a DRW model to obtain priors for the DRW parameters ($\tau_{\rm DRW}$, ${\rm SF}_{\rm inf}$), then fits all light curves for the shared DRW parameters and individual transfer function parameters for each light curve at longer wavelengths. The transfer function is modeled as a narrow top-hat function with three parameters, lag, width, and scale. The updated {\tt JAVELIN} includes a module for continuum RM \citep{Mudd_etal_2018}, which fits the best thin disk model using Equation \ref{eq:lag_disksize} instead of individual lags at each wavelength. Similarly, we use Equation \ref{eq:disksize} to calculate ${M}_{\rm BH}{\dot{M}}$ using the disk size fitted by {\tt JAVELIN}. We use 100 walkers, 500 burn-in steps, and 1000 steps for the MCMC fitting, and limit the lag range to $\pm30$\,days.

\subsubsection{{\tt CREAM}/{\tt pycecream}}
{\tt CREAM} \citep{Starkey_etal_2016} is another MCMC code developed for CRM. It models a driving light curve at a very short wavelength and the full response function (i.e., Equation \ref{eq:response}) simultaneously to constrain ${M}_{\rm BH}{\dot{M}}$ and $i$. The driving light curve in {\tt CREAM} is modeled by a high-order Fourier series, free of the DRW assumption. Since {\tt CREAM} fit for the full response function, instead of a top-hat transfer function, it is the only traditional method that can constrain the disk inclination. \cite{Chan_etal_2020} found CREAM is more accurate than other methods because it consider the more realistic, skewed transfer function. We run 4 independent chains for 10,000 steps and visually check the chains for convergence using the Python-implemented {\tt pycecream}\footnote{https://github.com/drds1/pycecream}.

\begin{figure*}
    \centering
    \includegraphics[width=\textwidth]{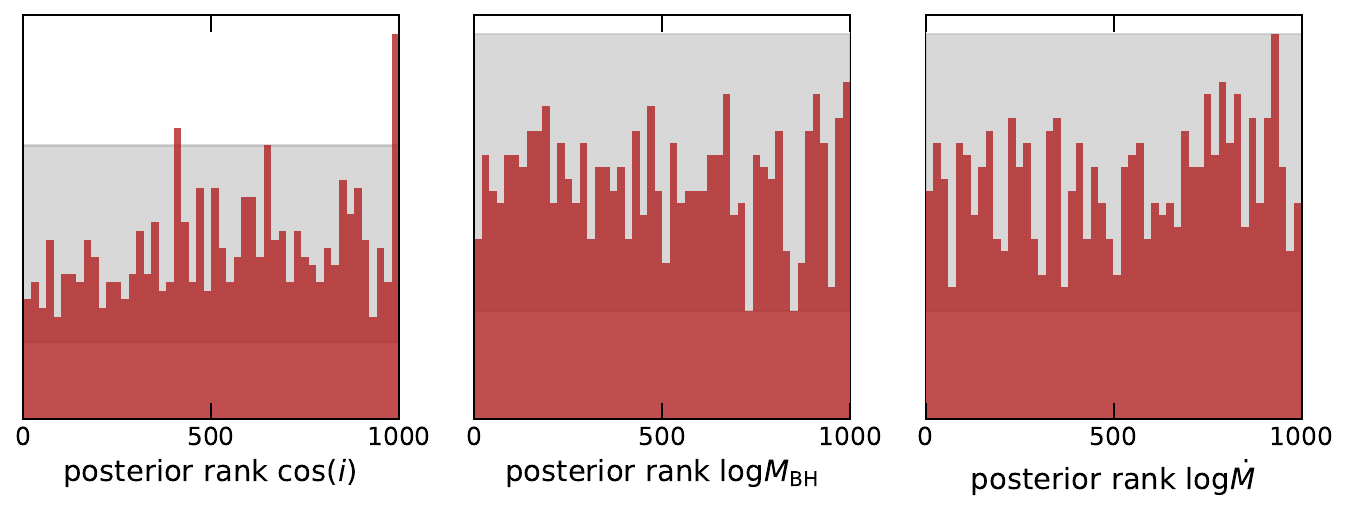}
    \caption{Rank distributions from SBC, from left to right panels are cos($i$), ${M}_{\rm BH}$, and ${\dot{M}}$. The grey shaded areas indicate the 99\% variance of a uniform distribution, i.e., a well-calibrated inference should only have one bar outside of the grey area. The SBC test indicates the posterior is robust for ${M}_{\rm BH}$ and ${\dot{M}}$ and slightly underestimated for cos($i$).}
    \label{fig:sbc}
\end{figure*}

\section{Results}\label{sec:result}

\subsection{Overall performance of the SBI model}\label{sec:overall_result}

{To evaluate the performance of the SBI model, we draw 1,000 samples from the approximate posterior distribution, which is built from the trained neural network, for each light curve in the testing data and take the median (and the 16, 84 percentiles) as the inferred parameter values (and their 1$\sigma$ uncertainties).}
Figure \ref{fig:sbi_result} shows the overall performance of the SBI model. Since the light curves are generated from \cite{MacLeod_etal_2012} empirical relations, ${ M}_{\rm BH}$ and ${\dot{M}}$ are not fully degenerate and can be individually constrained. The top row of Figure \ref{fig:sbi_result} shows the true and predicted BH parameters in the testing data, and the middle row shows the distribution of the deviation between the truth and prediction. Both ${M}_{\rm BH}$ and ${\dot{M}}$ are well-constrained by SBI, with a scatter of 0.08\,dex over the entire prior range. $cos(i)$ is constrained to within $\pm$0.01, which is roughly a few degrees at the prior range ($i=0-45 ^{\circ}$). Finally, the bottom row of Figure \ref{fig:sbi_result} shows the fraction between the deviation from the true value and the $1\sigma$ uncertainty from the posterior. All three curves show a good match to a Gaussian profile of unity, suggesting the uncertainties are reasonable.

For Bayesian inference algorithms, like MCMC or SBI, it is difficult to evaluate the algorithms' robustness without computing the {true} posterior, which is impossible to compute for all but very simple physical models. Simulation-based calibration \citep[SBC,][]{Cook_etal_2006, Talts_etal_2018} is a tool for validating the inference robustness by comparing the {\it average} posterior distribution of simulated data to the wide prior range that is used to generate the aforementioned simulated data. To perform SBC, we simulate 1,000 light curve sets using uniform priors across the entire prior range and evaluate the posterior of each light curve set. If the trained SBI model is robust, the composite posterior of all test cases should be same as the uniform prior distribution, as indicated by the Bayes' theorem. Since the SBI posterior is amortized, SBC is efficient and only limited by the computation resource needed for generating new simulated data. In Figure \ref{fig:sbc}, we show the overall rank distribution comparing the prior and posterior. The uniform distributions (within the uncertainty denoted in the grey band) suggest the posterior from the trained model is not biased or skewed for ${M}_{\rm BH}$ and ${\dot{M}}$. The cosine of inclination cos($i$) shows a slightly right-skewed rank distribution, suggesting there is a systematic underestimation of cos($i$), which is expected as cos($i$) cannot be over 1.

\begin{figure}
    \centering
    \includegraphics[width=0.45\textwidth]{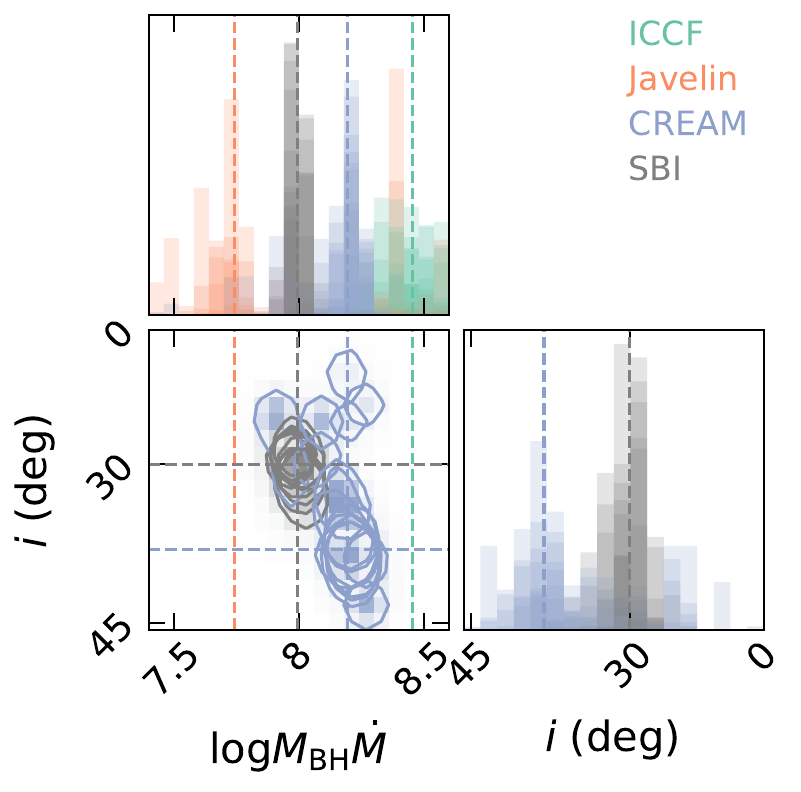}
    \caption{Posterior distribution of ${M}_{\rm BH}\dot{M}$ and $i$ from SBI and traditional inference methods. The histograms show results from 10 different sets of testing light curves with the same BH parameters, as described in Section \ref{sec:compare}. The dashed lines show the median of the combined posteriors.}
    \label{fig:compare}
\end{figure}

\subsection{Comparison with traditional inference methods}\label{sec:compare}
Due to the long computation time of {\tt JAVELIN} and {\tt CREAM}, we compare the SBI results with the traditional methods using 10 example light curve sets with ${M}_{\rm BH}=10^{8}\,{\rm M}_{\Sun}$, ${\dot{M}}=1\,{\rm M}_{\Sun}$\,yr$^{-1}$, and $i=30^{\circ}$. Since ${M}_{\rm BH}$ and ${\dot{M}}$ are degenerate for the thin-disk model (Equation \ref{eq:disksize}), we compare ${M}_{\rm BH}\dot{M}$ for all methods, and $i$ for SBI and {\tt CREAM}. 

Figure \ref{fig:compare} shows the ${M}_{\rm BH}\dot{M}$ and $i$ posterior distributions of all 10 example light curves from each method. We find that ICCF in general produces the least accurate results with large uncertainties, {\tt JAVELIN} tends to underestimate ${M}_{\rm BH}\dot{M}$, {and {\tt CREAM} is most accurate in constraining ${M}_{\rm BH}\dot{M}$ and $i$ in all example light curves among the three methods. There is a small offset in the estimated ${M}_{\rm BH}\dot{M}$ and $i$ between {\tt CREAM} and the input value due to small differences in simulation parameters and numerical calculations.} These results are in agreement with previous studies on lag methodologies. \cite{Jiang_etal_2017} and \cite{Yu_etal_2020} showed ICCF is more likely to fail at measuring lags or produce large uncertainties compared to {\tt JAVELIN}, while {\tt JAVELIN} can recover lags smaller than the observing cadence by implementing the more realistic DRW interpolation \citep{Li_etal_2019}. However, {\tt JAVELIN} only considers top-hat transfer functions, which is insufficient for high-cadence light curves for CRM \citep{Chan_etal_2020}. \cite{Chan_etal_2020} showed {\tt CREAM} is the most accurate of all traditional methods as it considers the full transfer function (Equation \ref{eq:response}). In this work, we simulate the training light curves with the full transfer function, so our SBI model can accurately constrain ${M}_{\rm BH}\dot{M}$ as if we fully modeled the skewed transfer function.

\begin{figure*}
    \centering
    \includegraphics[width=\textwidth]{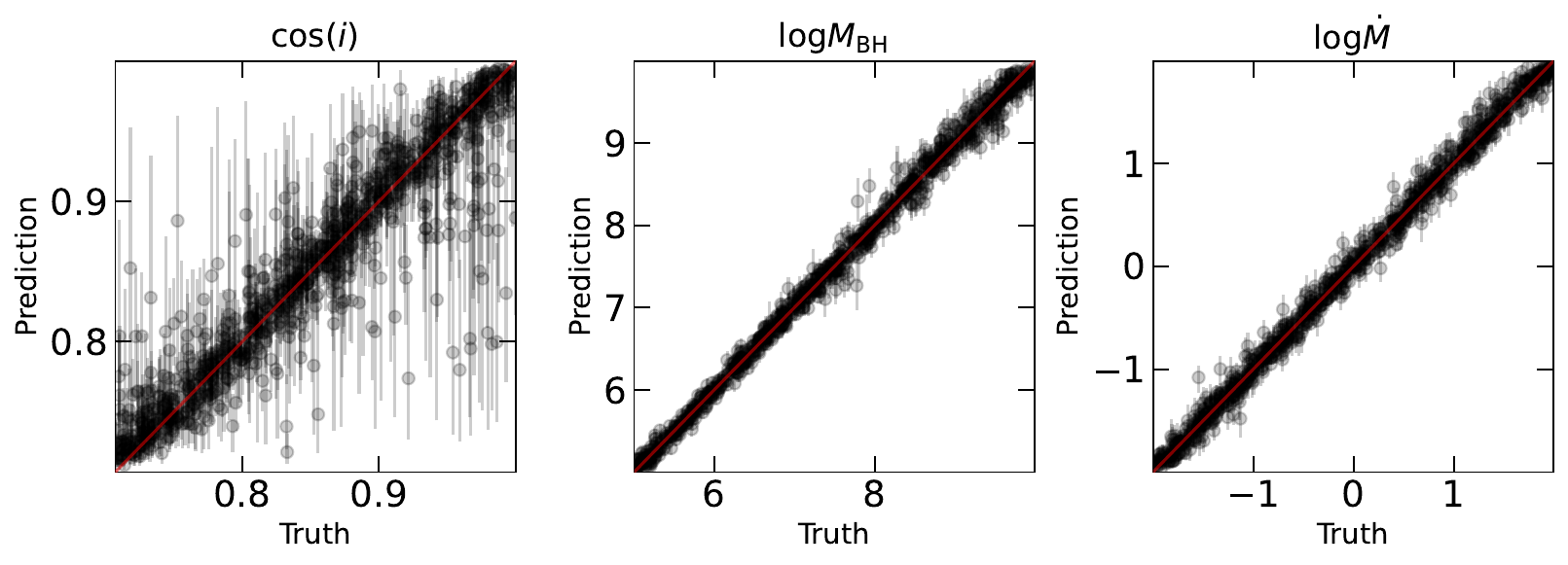}\\    
    \includegraphics[width=\textwidth]{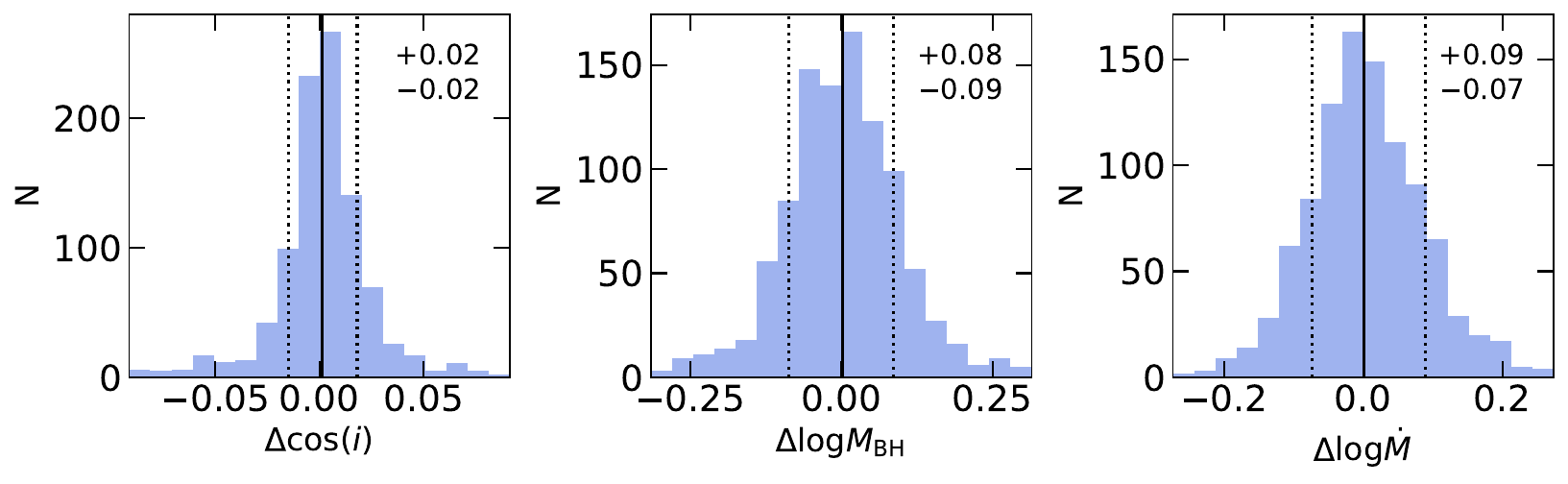}\\    
    \includegraphics[width=\textwidth]{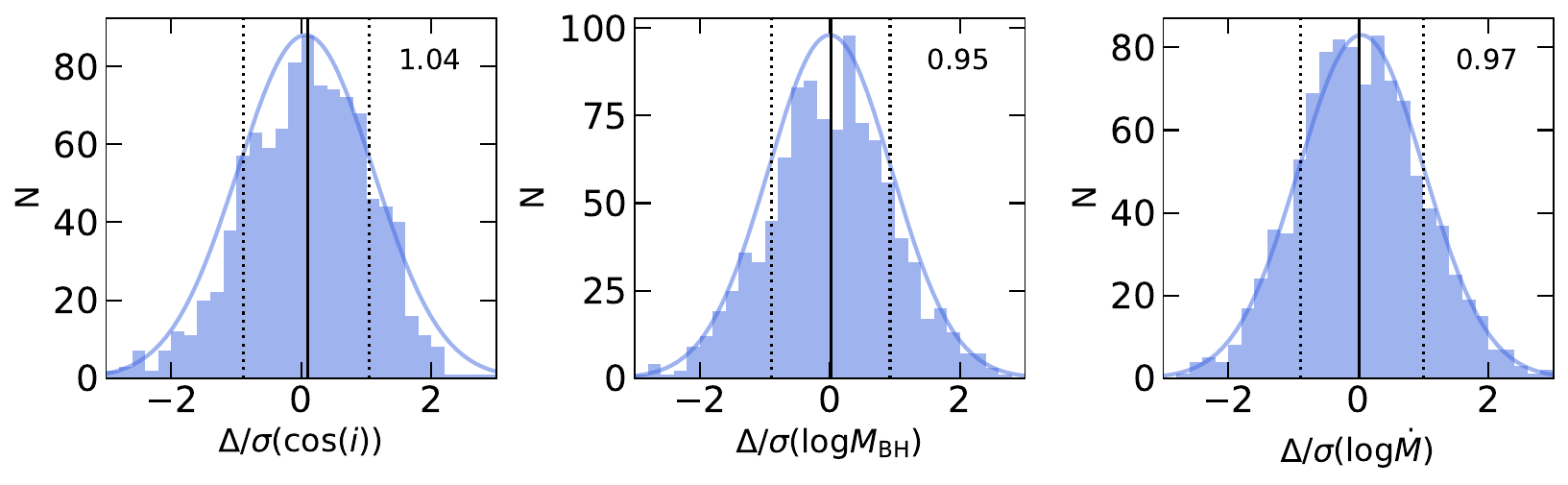}
    \caption{Evaluation results of testing data with large filter gaps (Section \ref{sec:discuss_gap}). Figure description is the same as Figure \ref{fig:sbi_result}. The trained SBI model can constrain $M_{\rm BH}$ and $\dot{M}$ to $\sim$0.1\,dex, and $i$ to $\lesssim10$ degrees.}
    \label{fig:sbi_result_gap}
\end{figure*}

\subsection{Computation Time}
Each DDF in LSST is expected to monitor thousands of AGN.  Therefore, efficient inference methods are extremely important to fully realize the scientific potential of LSST. Here, we provide a rough comparison of the computation resources needed for each method. 

For ICCF, the majority of computation time is spent on drawing MC realizations to estimate lag uncertainties. Typically, calculating lags between a few photometric bands with 500 realizations will take $\sim$minutes running on standard personal computers (e.g., Macbook Pro with M1 chip, 8 cores, and 16 GB memory). The computation can be easily parallelized for large number of targets on multiple computers or computing clusters, so ICCF is still computationally efficient for handling LSST DDF data. However, for {\tt JAVELIN} and {\tt CREAM}, the MCMC computation may take several hours to converge for each target, and therefore is no longer feasible for future CRM surveys. 

For the SBI setup in this paper, the majority of computation time is spent generating the simulated light curves. It took 100 CPU-hours to generate $10^5$ simulations for this work. 
The neural network training took roughly 6 hours for $10^5$ simulations on 1 GPU (NVIDIA Tesla V100) on the University of Michigan Great Lakes Cluster. The total evaluation time for the testing data (1000 light curve sets) is $\sim$30 second on 1 GPU (or $\sim$150 seconds on a 36-core CPU machine), which is a few minutes for $\sim3000$ quasars in a single DDF. {The inference speed is $\sim10^{3}$ times faster than ICCF and $\sim10^{5}$ times faster than {\tt Javelin} and {\tt CREAM}.}


\section{Discussion}\label{sec:discussion}

\subsection{Filter gaps and interpolation}\label{sec:discuss_gap}

Time-domain surveys in Astronomy usually have sparse and irregular time sampling due to survey design, bad weather, and technical difficulties. How to interpolate light curves and avoid artificial signals from the observing cadence pattern have always been the main issues in RM methodologies. ICCF linearly interpolates light curves onto a evenly-spaced grid before calculating the cross correlation at each time lag, and the standard RSS procedure randomly select subsets of data points to mitigate the effects of light curve sampling. {\tt JAVELIN} and {\tt CREAM} use the DRW model and high-order Fourier series to model the full driving light curve at high cadences (e.g., 0.1 days) jointly with the BH parameters. Additionally, the number of overlapping data points and the auto-correlation of light curves are used to evaluate if the lags are likely real or aliases in practice \citep[e.g.,][]{Grier_etal_2017, Grier_etal_2019, Homayouni_etal_2020}.

A major limitation of ML is that the training data has to have similar characteristics to the observed data in order to provide good results. Here, we take a different approach to deal with sparse, irregularly-sampled light curves. Since RM analysis is usually done after each observing season, the observed baseline will be known and can be incorporated to the simulated training data. This is particularly convenient for wide-field surveys like LSST as thousands of light curve sets from a single pointing will have identical observing patterns.

While the DDF observing strategy has not been finalized, the Phase 2 recommendations\footnote{https://pstn-055.lsst.io/} from the Survey Cadence Optimization Committee (SOCC) recommend the DDF program observe the $u$, $g$, $r$, $i$, $z$, and $Y$ filters in sequences. Since the filter wheel can fit only 5 filters at a time, observations of $u$ and $z$ band will alternate, $u$ band will be observed when lunar illumination is below 40\% and $z$ band will be available for the rest of the time. The details of different cadence, depth, filter swapping procedures are still being evaluated and could change in the future.

In this section, we train a new SBI model using simulated light curves with large gaps in $u$ and $z$ band following the SOCC recommendations. We use the same training set as in Section \ref{sec:result} but assign the flux value to -99 in the data gaps, which are identical for all targets. We train the SBI model using the same neural network setup and hyperparameters described in Section \ref{sec:method}, and evaluate on the 1,000 testing light curves after applying the same data gaps. Figure \ref{fig:sbi_result_gap} shows the results of the comparison between the prediction, difference from truth, and uncertainty level. The SBI model trained with missing data performs well over the entire prior range, but the uncertainties are larger than the idealized benchmark trained without data gaps; ${M}_{\rm BH}$ and $\dot{M}$ are constrained to 0.1\,dex, and $i$ is constrained to $<10^{\circ}$. The computation time for training the {\tt sbi} model and evaluating the test data are on the same order as the original runs on the full light curves. 
This simplified test showed that {\tt sbi} models can yield good parameter estimations for light curves with large data gaps. We plan to investigate how different cadence strategy could affect RM results in future work, including incorporating realistic flux uncertainties and priors on AGN properties (e.g., redshift, luminosity function, etc...).  

Another approach to deal with missing data in SBI is to marginalize over possible realizations of the missing data to create ensemble posteriors \citep{Wang_etal_2023}. For example, one can interpolate the missing data points with DRW or other stochastic models and evaluate the posterior multiple times with the trained model. Since the posterior is amortized, the computation is still more efficient than traditional MCMC methods. Recently-developed Python packages like {\tt celerite} \citep{celerite} and {\tt tinygp} \citep{tinygp} also provide fast Gaussian Processes (GP) fitting framework for fitting AGN light curves with DRW or other stochastic models. However, we find that this approach does not work with large data gaps (e.g., 10-20 day gaps in this case), since there are too much missing data and insufficient constraints from GP fitting to match the variability signatures of the original light curves.

\begin{figure*}
    \centering
    \includegraphics[width=\textwidth]{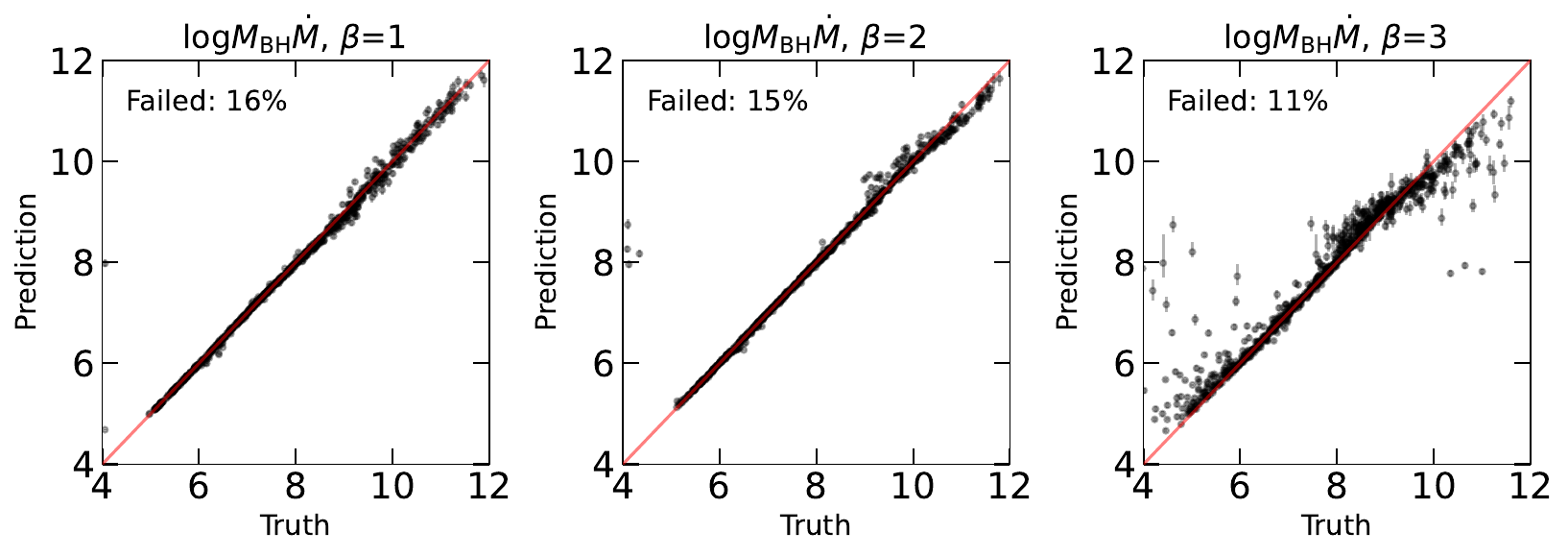}
    \caption{Evaluation results of the power-law PSD light curves with $\beta=1,2,3$ (left to right). The solid red lines denote the 1:1 relation. The fraction of failed evaluations are labeled in the top left corner of each panel.}
    \label{fig:nonDRW}
\end{figure*} 

\begin{figure*}
    \centering
    \includegraphics[width=\textwidth]{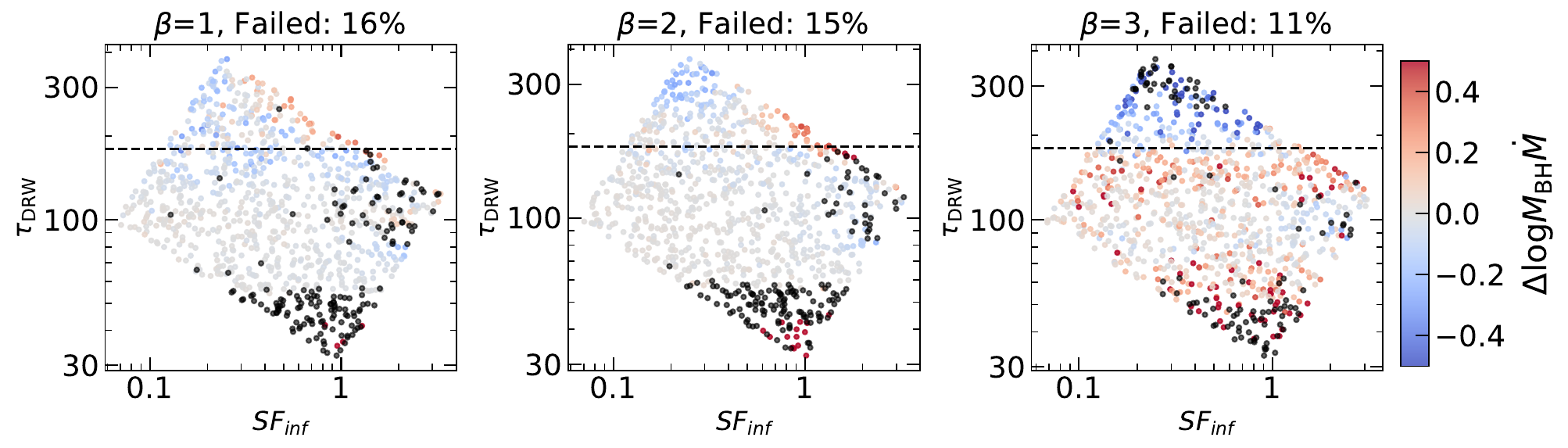}
    \caption{Accuracy of the ${M}_{\rm BH}\dot{M}$ prediction as functions of the ``expected'' DRW parameters. The color of each point shows the deviation of ${M}_{\rm BH}\dot{M}$ prediction from the truth, and the black points indicate the failed evaluation. {The accuracy of the prediction is correlated with the expected DRW damping timescale, suggesting the neural network has extracted features related to the damping timescale.}}
    \label{fig:nonDRW_tau}
\end{figure*}

\subsection{Deviation from DRW}\label{sec:discuss_drw}

While the DRW model is generally a good approximation for light curves at days-to-weeks timescales, it is not clear if AGN variability are DRW at smaller or longer timescales \citep{Kasliwal_etal_2015, Moreno_etal_2021, Stone_etal_2022}. In addition, other physical processes (e.g., SMBH binaries, tidal disruption events) may produce additional variability on top of the intrinsic AGN variability. 

In this section, we test the performance of our original trained SBI model from Section \ref{sec:overall_result} on non-DRW light curves. We generate three sets of daily-sampled testing light curves where the driving light curves have power-law spectral density function (PSD) ${\rm P}\propto(1/f)^\beta$ of varying slopes ($\beta=1, 2, 3$). The DRW model has a broken power-law PSD, with slope of 2 at high frequency and 0 at low frequency. 

Figure \ref{fig:nonDRW} shows the predicted ${M}_{\rm BH}\dot{M}$ when evaluating the non-DRW light curves with the model trained on DRW light curves. For the $\beta=1, 2$ light curves, ${M}_{\rm BH}\dot{M}$ can be {constrained to within $\sim0.08$\,dex and $\sim0.05$\,dex over most of the prior range, though the prediction accuracy is lower than the DRW light curves (1$\sigma\sim0.02$\,dex). For $\beta= 3$ light curves, the ${M}_{\rm BH}\dot{M}$ prediction is {less robust (1$\sigma\sim0.2$\,dex), with more outliers in the high and low ${M}_{\rm BH}\dot{M}$ ends} and no obvious trends in the deviation from the truth.}  This is not solely a problem for SBI. In traditional MCMC methods like {\tt JAVELIN}, BH parameters predicted from non-DRW light curves are also less accurate, as shown in \cite{Li_etal_2019}.

In Figure \ref{fig:nonDRW_tau} (especially for $\beta= 3$ light curves), we show that the difference between the prediction and truth is correlated to the ``expected" damping timescale of the BH parameters. If the variability signature of the non-DRW light curves are similar to the DRW light curves with similar ``expected" damping timescale, the trained SBI model can infer the BH parameters by treating the non-DRW light curves as DRW light curves. Otherwise, the behavior of the SBI model is hard to predict. Interestingly, the prediction accuracy does not correlate with the ``expected'' variability magnitude, suggesting the SBI model has extracted and utilized features correlated with the damping timescale, but not the variability magnitude. 

{When the evaluated light curves are too far from the DRW light curves seen by the trained model, acceptance rate for the posterior sampling could decrease as samples lying outside of the prior bounds are rejected. For a small fraction of the non-DRW test cases, more than {$\>$50\%} of the samples are rejected, which we label as failed evaluations. Compared to $<1\%$ failed for the idealized testing data, roughly 16, 15, 11\% test cases failed for the $\beta=1, 2, 3$ light curves.}

For real observations, it is difficult to obtain the ground truth for calibrating how well the SBI model perform in different parameter ranges. {Therefore, pre-trained SBI models should not be applied to out-of-distribution observed data. Nonetheless, if the simple pre-trained models failed at estimating the posteriors for many light curves, it would indicate that AGN light curves might deviate DRW models and alternative AGN variability model need to be considered.}
Alternatively, we can include more variability models in the training data to ensure the SBI model has been trained on a variety of variability characteristic.

\subsection{A Flexible Framework for CRM}

The key to successfully applying SBI to real data is building a simulation that is as realistic as possible. In this work, we adopted the simplest setup for the accretion disk geometry, reprocessing mechanism, and variability characteristics. However, previous wide-field and intensive CRM campaigns have shown observational evidence that points to more complicated physical picture of the innermost AGN. While accretion disk profiles typically follow the wavelength-lag relations predicted by the standard optically-thick, geometrically-thin accretion disk model, the size of accretion disks are still uncertain. \cite{Jiang_etal_2017} and other intensive CRM programs \citep{Edelson_etal_2015, Fausnaugh_etal_2016} have found accretion disks are typically 2-3 times larger than predicted, whereas the SDSS-RM and OzDES surveys reported accretion disk sizes consistent with the standard model \citep{Homayouni_etal_2019, Yu_etal_2020}. In addition, intensive RM programs revealed poor correlation between X-ray and UV/optical variability, suggesting X-ray emission might not the be main driver of the UV/optical light curves \citep{Starkey_etal_2017, McHardy_etal_2018, HernandezSantisteban_etal_2020}. Some observations can be explained by two-step reprocessing model \citep{Edelson_etal_2017, Gardner_Done_2017, Akiba_etal_2023}, heavily obscured geometry of the X-ray emitting region \citep{Cackett_etal_2023}, or complicated dynamics and gas flows in the accretion disk \citep{Starkey_etal_2023, Zaidouni_etal_2024}. There could also be additional sources of variability that are not associated to RM, for example, temperature fluctuations internal to the accretion disk \citep{HernandezSantisteban_etal_2020, Neustadt_etal_2022, Stone_etal_2023, Neustadt_etal_2024} and SMBH binaries \citep{Chen_etal_2020, Liao_etal_2021}.

As the SBI method only requires simulated observations (e.g., light curves) and their corresponding physical parameters (e.g., BH parameters) as inputs, it can be a flexible framework to include alternative physical assumptions, e.g., accretion disk geometry, reprocessing mechanisms, and variability characteristics, {or draw inference from numerical simulations \citep[e.g.,][]{Secunda_etal_2023}}. A general model can include training data with various assumptions to improve parameter estimation over the general AGN population, alternatively, multiple models used to evaluate how well can different assumptions explain the observed AGN light curves.

\section{Conclusions}\label{sec:conclusion}
Using simulated light curves, we demonstrate that SBI can be an efficient method for estimating BH parameters in future CRM surveys. In our idealized test case without missing data, the trained SBI model can constrain ${M}_{\rm BH}$ and ${\dot{M}}$ to within $\pm$0.08\,dex and $i$ to a few degrees, which is as good as {\tt CREAM} and more accurate than ICCF and {\tt JAVELIN}. Due to the nature of amortized posteriors, SBI is 100x faster than ICCF and 10,000x faster than MCMC methods. {The SBI framework is particularly efficient for wide-field RM surveys, as  the observing strategy can be incorporated into SBI simulations to train a general SBI model for all light curves.}

We also explored more realistic test cases. By incorporating the planned observing baseline into the SBI training, we find BH parameters can be constrained better than using traditional methods for data with large filter gaps. {When the assumptions of the training data are violated (e.g., AGN variability is not DRW), the posterior sampling could be less accurate or rejected. If a simple SBI model failed, it would imply different physical model for AGN variability and accretion disk geometry.} In the future, we plan to implement realistic observation strategy and uncertainty in our SBI model, as well as incorporating more sophisticated variability and accretion disk models, with the goal of building an SBI framework suited for analyzing LSST data. {The SBI framework can also be extended to other RM surveys, including broad-line region RM and torus RM surveys, to compare simulation and observed light curves directly.}

\begin{acknowledgments}
JIL is supported by the Eric and Wendy Schmidt AI in Science Postdoctoral Fellowship, a Schmidt Futures program. CA acknowledges support from the Leinweber Center for Theoretical Physics and DOE grant DE-SC009193.  YS acknowledges partial support from NSF grant AST-2009947. 
\end{acknowledgments}

\software{{\tt CREAM} \citep{Starkey_etal_2016}, {\tt JAVELIN} \citep{Zu_etal_2013,Mudd_etal_2018}, {\tt PyCCF} \citep{PyCCF}, {\tt sbi}\citep{sbi}}

\appendix
\section{Additional SBI configurations}\label{app:other_method}
{We experiment with different setups of the embedding summary network and list the evaluation results here. We use the embedding summary networks to extract features from the simulated light curves (180$\times$7), then the features are fed into the MAF neural density estimator (described in Section \ref{sec:model}) to build the approximate posterior. We find that using embedding summary networks can greatly improve the inference accuracy, and LSTM networks outperform a basic convolution neural network (CNN) due to its ability to extract time-dependent variability features.}

\restartappendixnumbering 

\begin{table*}[h]
\hspace{-1cm}
    \begin{tabular}{lllrrrr}
    \hline\hline
      Name  &  Layer  & Output Shape & \# of Parameters & $\sigma$(cos($i$)) & $\sigma$($\log{M}_{\rm BH}$) & $\sigma$($\log{\dot{M}}$) \\
    \hline
    Defualt SBI & N/A & (Batch, 1260)  & -- & 0.10    & 0.33 & 0.23 \\
    \hline
    {\bf LSTM} &  &   & (19,728) & {\bf 0.01}    & {\bf 0.08} & {\bf 0.08} \\
    {\bf (last timestep)} & LSTM & (Batch, 1, 64)  & 18,688 &  & & \\
    & Linear & (Batch, 16)  & 1,040 &    &  & \\
    \hline    
    LSTM  &  &   & (76,336) & 0.01    & 0.08 & 0.08 \\
    ({\tt optuna}) & LSTM & (Batch, 1, 128)  & 70,144 &   &  & \\
    & Linear & (Batch, 48)  & 6,192 &    &  & \\
    \hline    
    LSTM  &  &   & (203,024) & 0.01    & 0.09 & 0.08 \\
    (all timesteps) & LSTM & (Batch, 180, 64)  & 18,688 &   &  & \\
    & Linear & (Batch, 16)  & 184,336 &    &  & \\
    \hline    
    CNN &  &   & (11,692) & 0.06    & 0.14 & 0.09 \\
    & Conv2D & (Batch, 6, 180, 7)  & 156 &   &  & \\
    & MaxPool2D & (Batch, 6, 60, 2)  & -- &    &  & \\
    & Linear & (Batch, 16)  & 11,536 &    &  & \\
    \hline\hline
    \end{tabular}
    \caption{Architectures and results from different SBI configurations. The numbers in brackets are the total number of hidden parameters in each summary network. }
    \label{tab:SBIparameters}
\end{table*}

\bibliography{refs} 

\begin{thebibliography}{}
\expandafter\ifx\csname natexlab\endcsname\relax\def\natexlab#1{#1}\fi
\providecommand{\url}[1]{\href{#1}{#1}}
\providecommand{\dodoi}[1]{doi:~\href{http://doi.org/#1}{\nolinkurl{#1}}}
\providecommand{\doeprint}[1]{\href{http://ascl.net/#1}{\nolinkurl{http://ascl.net/#1}}}
\providecommand{\doarXiv}[1]{\href{https://arxiv.org/abs/#1}{\nolinkurl{https://arxiv.org/abs/#1}}}

\bibitem[{{Akiba} {et~al.}(2023){Akiba}, {Dexter}, {Brandt}, {Ho}, {Homayouni},
  {Schneider}, {Shen}, \& {Trump}}]{Akiba_etal_2023}
{Akiba}, T., {Dexter}, J., {Brandt}, W.~N., {et~al.} 2023, \apj, 953, 124,
  \dodoi{10.3847/1538-4357/ace1e1}

\bibitem[{Akiba {et~al.}(2019)Akiba, Sano, Yanase, Ohta, \& Koyama}]{optuna}
Akiba, T., Sano, S., Yanase, T., Ohta, T., \& Koyama, M. 2019, in The 25th ACM
  SIGKDD International Conference on Knowledge Discovery \& Data Mining,
  2623--2631

\bibitem[{{Brandt} {et~al.}(2018){Brandt}, {Ni}, {Yang}, {Anderson}, {Assef},
  {Barth}, {Bauer}, {Bongiorno}, {Chen}, {De Cicco}, {Gezari}, {Grier}, {Hall},
  {Hoenig}, {Lacy}, {Li}, {Luo}, {Paolillo}, {Peterson}, {Popovi{\'c}},
  {Richards}, {Shemmer}, {Shen}, {Sun}, {Timlin}, {Trump}, {Vito}, \&
  {Yu}}]{Brandt_etal_2018}
{Brandt}, W.~N., {Ni}, Q., {Yang}, G., {et~al.} 2018, arXiv e-prints,
  arXiv:1811.06542, \dodoi{10.48550/arXiv.1811.06542}

\bibitem[{{Cackett} {et~al.}(2021){Cackett}, {Bentz}, \&
  {Kara}}]{Cackett_etal_2021}
{Cackett}, E.~M., {Bentz}, M.~C., \& {Kara}, E. 2021, iScience, 24, 102557,
  \dodoi{10.1016/j.isci.2021.102557}

\bibitem[{{Cackett} {et~al.}(2023){Cackett}, {Gelbord}, {Barth}, {De Rosa},
  {Edelson}, {Goad}, {Homayouni}, {Horne}, {Kara}, {Kriss}, {Korista}, {Landt},
  {Plesha}, {Arav}, {Bentz}, {Boizelle}, {Dalla Bont{\`a}}, {Dehghanian},
  {Donnan}, {Du}, {Ferland}, {Fian}, {Filippenko}, {Gonz{\'a}lez Buitrago},
  {Grier}, {Hall}, {Hu}, {Ili{\'c}}, {Kaastra}, {Kaspi}, {Kochanek},
  {Kova{\v{c}}evi{\'c}}, {Kynoch}, {Li}, {McLane}, {Mehdipour}, {Miller},
  {Montano}, {Netzer}, {Panagiotou}, {Partington}, {{\v{C}}. Popovi{\'c}},
  {Proga}, {Rogantini}, {Sanmartim}, {Siebert}, {Storchi-Bergmann},
  {Vestergaard}, {Wang}, {Waters}, \& {Zaidouni}}]{Cackett_etal_2023}
{Cackett}, E.~M., {Gelbord}, J., {Barth}, A.~J., {et~al.} 2023, \apj, 958, 195,
  \dodoi{10.3847/1538-4357/acfdac}

\bibitem[{{Chan} {et~al.}(2020){Chan}, {Millon}, {Bonvin}, \&
  {Courbin}}]{Chan_etal_2020}
{Chan}, J.~H.~H., {Millon}, M., {Bonvin}, V., \& {Courbin}, F. 2020, \aap, 636,
  A52, \dodoi{10.1051/0004-6361/201935423}

\bibitem[{{Chen} {et~al.}(2020){Chen}, {Liu}, {Liao}, {Holgado}, {Guo},
  {Gruendl}, {Morganson}, {Shen}, {Zhang}, {Abbott}, {Aguena}, {Allam},
  {Avila}, {Bertin}, {Bhargava}, {Brooks}, {Burke}, {Carnero Rosell},
  {Carollo}, {Carrasco Kind}, {Carretero}, {Costanzi}, {da Costa}, {Davis}, {De
  Vicente}, {Desai}, {Diehl}, {Doel}, {Everett}, {Flaugher}, {Friedel},
  {Frieman}, {Garc{\'\i}a-Bellido}, {Gaztanaga}, {Glazebrook}, {Gruen},
  {Gutierrez}, {Hinton}, {Hollowood}, {James}, {Kim}, {Kuehn}, {Kuropatkin},
  {Lewis}, {Lidman}, {Lima}, {Maia}, {March}, {Marshall}, {Menanteau},
  {Miquel}, {Palmese}, {Paz-Chinch{\'o}n}, {Plazas}, {Sanchez}, {Schubnell},
  {Serrano}, {Sevilla-Noarbe}, {Smith}, {Suchyta}, {Swanson}, {Tarle},
  {Tucker}, {Norbert Varga}, \& {Walker}}]{Chen_etal_2020}
{Chen}, Y.-C., {Liu}, X., {Liao}, W.-T., {et~al.} 2020, \mnras, 499, 2245,
  \dodoi{10.1093/mnras/staa2957}

\bibitem[{Cook {et~al.}(2006)Cook, Gelman, \& Rubin}]{Cook_etal_2006}
Cook, S.~R., Gelman, A., \& Rubin, D.~B. 2006, Journal of Computational and
  Graphical Statistics, 15, 675

\bibitem[{Cranmer {et~al.}(2020)Cranmer, Brehmer, \&
  Louppe}]{Cranmer_etal_2020}
Cranmer, K., Brehmer, J., \& Louppe, G. 2020, Proceedings of the National
  Academy of Sciences, 117, 30055

\bibitem[{{De Rosa} {et~al.}(2015){De Rosa}, {Peterson}, {Ely}, {Kriss},
  {Crenshaw}, {Horne}, {Korista}, {Netzer}, {Pogge}, {Ar{\'e}valo}, {Barth},
  {Bentz}, {Brandt}, {Breeveld}, {Brewer}, {Dalla Bont{\`a}}, {De
  Lorenzo-C{\'a}ceres}, {Denney}, {Dietrich}, {Edelson}, {Evans}, {Fausnaugh},
  {Gehrels}, {Gelbord}, {Goad}, {Grier}, {Grupe}, {Hall}, {Kaastra}, {Kelly},
  {Kennea}, {Kochanek}, {Lira}, {Mathur}, {McHardy}, {Nousek}, {Pancoast},
  {Papadakis}, {Pei}, {Schimoia}, {Siegel}, {Starkey}, {Treu}, {Uttley},
  {Vaughan}, {Vestergaard}, {Villforth}, {Yan}, {Young}, \&
  {Zu}}]{DeRosa_etal_2015}
{De Rosa}, G., {Peterson}, B.~M., {Ely}, J., {et~al.} 2015, \apj, 806, 128,
  \dodoi{10.1088/0004-637X/806/1/128}

\bibitem[{{Edelson} {et~al.}(2015){Edelson}, {Gelbord}, {Horne}, {McHardy},
  {Peterson}, {Ar{\'e}valo}, {Breeveld}, {De Rosa}, {Evans}, {Goad}, {Kriss},
  {Brandt}, {Gehrels}, {Grupe}, {Kennea}, {Kochanek}, {Nousek}, {Papadakis},
  {Siegel}, {Starkey}, {Uttley}, {Vaughan}, {Young}, {Barth}, {Bentz},
  {Brewer}, {Crenshaw}, {Dalla Bont{\`a}}, {De Lorenzo-C{\'a}ceres}, {Denney},
  {Dietrich}, {Ely}, {Fausnaugh}, {Grier}, {Hall}, {Kaastra}, {Kelly},
  {Korista}, {Lira}, {Mathur}, {Netzer}, {Pancoast}, {Pei}, {Pogge},
  {Schimoia}, {Treu}, {Vestergaard}, {Villforth}, {Yan}, \&
  {Zu}}]{Edelson_etal_2015}
{Edelson}, R., {Gelbord}, J.~M., {Horne}, K., {et~al.} 2015, \apj, 806, 129,
  \dodoi{10.1088/0004-637X/806/1/129}

\bibitem[{{Edelson} {et~al.}(2017){Edelson}, {Gelbord}, {Cackett}, {Connolly},
  {Done}, {Fausnaugh}, {Gardner}, {Gehrels}, {Goad}, {Horne}, {McHardy},
  {Peterson}, {Vaughan}, {Vestergaard}, {Breeveld}, {Barth}, {Bentz},
  {Bottorff}, {Brandt}, {Crawford}, {Dalla Bont{\`a}}, {Emmanoulopoulos},
  {Evans}, {Figuera Jaimes}, {Filippenko}, {Ferland}, {Grupe}, {Joner},
  {Kennea}, {Korista}, {Krimm}, {Kriss}, {Leonard}, {Mathur}, {Netzer},
  {Nousek}, {Page}, {Romero-Colmenero}, {Siegel}, {Starkey}, {Treu}, {Vogler},
  {Winkler}, \& {Zheng}}]{Edelson_etal_2017}
{Edelson}, R., {Gelbord}, J., {Cackett}, E., {et~al.} 2017, \apj, 840, 41,
  \dodoi{10.3847/1538-4357/aa6890}

\bibitem[{{Edelson} {et~al.}(2019){Edelson}, {Gelbord}, {Cackett}, {Peterson},
  {Horne}, {Barth}, {Starkey}, {Bentz}, {Brandt}, {Goad}, {Joner}, {Korista},
  {Netzer}, {Page}, {Uttley}, {Vaughan}, {Breeveld}, {Cenko}, {Done}, {Evans},
  {Fausnaugh}, {Ferland}, {Gonzalez-Buitrago}, {Gropp}, {Grupe}, {Kaastra},
  {Kennea}, {Kriss}, {Mathur}, {Mehdipour}, {Mudd}, {Nousek}, {Schmidt},
  {Vestergaard}, \& {Villforth}}]{Edelson_etal_2019}
---. 2019, \apj, 870, 123, \dodoi{10.3847/1538-4357/aaf3b4}

\bibitem[{{Fagin} {et~al.}(2024){Fagin}, {Park}, {Best}, {Chan}, {Ford},
  {Graham}, {Villar}, {Ho}, \& {O'Dowd}}]{Fagin_etal_2024}
{Fagin}, J., {Park}, J.~W., {Best}, H., {et~al.} 2024, \apj, 965, 104,
  \dodoi{10.3847/1538-4357/ad2988}

\bibitem[{{Fausnaugh} {et~al.}(2016){Fausnaugh}, {Denney}, {Barth}, {Bentz},
  {Bottorff}, {Carini}, {Croxall}, {De Rosa}, {Goad}, {Horne}, {Joner},
  {Kaspi}, {Kim}, {Klimanov}, {Kochanek}, {Leonard}, {Netzer}, {Peterson},
  {Schn{\"u}lle}, {Sergeev}, {Vestergaard}, {Zheng}, {Zu}, {Anderson},
  {Ar{\'e}valo}, {Bazhaw}, {Borman}, {Boroson}, {Brandt}, {Breeveld}, {Brewer},
  {Cackett}, {Crenshaw}, {Dalla Bont{\`a}}, {De Lorenzo-C{\'a}ceres},
  {Dietrich}, {Edelson}, {Efimova}, {Ely}, {Evans}, {Filippenko}, {Flatland},
  {Gehrels}, {Geier}, {Gelbord}, {Gonzalez}, {Gorjian}, {Grier}, {Grupe},
  {Hall}, {Hicks}, {Horenstein}, {Hutchison}, {Im}, {Jensen}, {Jones},
  {Kaastra}, {Kelly}, {Kennea}, {Kim}, {Korista}, {Kriss}, {Lee}, {Lira},
  {MacInnis}, {Manne-Nicholas}, {Mathur}, {McHardy}, {Montouri}, {Musso},
  {Nazarov}, {Norris}, {Nousek}, {Okhmat}, {Pancoast}, {Papadakis}, {Parks},
  {Pei}, {Pogge}, {Pott}, {Rafter}, {Rix}, {Saylor}, {Schimoia}, {Siegel},
  {Spencer}, {Starkey}, {Sung}, {Teems}, {Treu}, {Turner}, {Uttley},
  {Villforth}, {Weiss}, {Woo}, {Yan}, \& {Young}}]{Fausnaugh_etal_2016}
{Fausnaugh}, M.~M., {Denney}, K.~D., {Barth}, A.~J., {et~al.} 2016, \apj, 821,
  56, \dodoi{10.3847/0004-637X/821/1/56}

\bibitem[{{Foreman-Mackey} {et~al.}(2020){Foreman-Mackey}, {Agol}, {Angus},
  {Brewer}, {Austin}, {Casey}, {Czekala}, {Guillochon}, \& {Meierjurgen
  Farr}}]{celerite}
{Foreman-Mackey}, D., {Agol}, E., {Angus}, R., {et~al.} 2020, {dfm/celerite:
  celerite v0.4.0}, v0.4.0,  Zenodo, \dodoi{10.5281/zenodo.3934421}

\bibitem[{{Foreman-Mackey} {et~al.}(2024){Foreman-Mackey}, {Yu}, {Yadav},
  {Reynolds Becker}, {Caplar}, {Huppenkothen}, {Killestein}, {Tronsgaard},
  {Rashid}, \& {Schmerler}}]{tinygp}
{Foreman-Mackey}, D., {Yu}, W., {Yadav}, S., {et~al.} 2024, {dfm/tinygp: The
  tiniest of Gaussian Process libraries}, v0.3.0,  Zenodo,
  \dodoi{10.5281/zenodo.10463641}

\bibitem[{{Frank} {et~al.}(2002){Frank}, {King}, \& {Raine}}]{Frank_etal_2002}
{Frank}, J., {King}, A., \& {Raine}, D.~J. 2002, {Accretion Power in
  Astrophysics: Third Edition}

\bibitem[{{Gardner} \& {Done}(2017)}]{Gardner_Done_2017}
{Gardner}, E., \& {Done}, C. 2017, \mnras, 470, 3591,
  \dodoi{10.1093/mnras/stx946}

\bibitem[{Greenberg {et~al.}(2019)Greenberg, Nonnenmacher, \& Macke}]{SNPE}
Greenberg, D., Nonnenmacher, M., \& Macke, J. 2019, in Proceedings of Machine
  Learning Research, Vol.~97, Proceedings of the 36th International Conference
  on Machine Learning, ed. K.~Chaudhuri \& R.~Salakhutdinov (PMLR), 2404--2414.
\newblock \url{https://proceedings.mlr.press/v97/greenberg19a.html}

\bibitem[{{Grier} {et~al.}(2017){Grier}, {Trump}, {Shen}, {Horne}, {Kinemuchi},
  {McGreer}, {Starkey}, {Brandt}, {Hall}, {Kochanek}, {Chen}, {Denney},
  {Greene}, {Ho}, {Homayouni}, {I-Hsiu Li}, {Pei}, {Peterson}, {Petitjean},
  {Schneider}, {Sun}, {AlSayyad}, {Bizyaev}, {Brinkmann}, {Brownstein},
  {Bundy}, {Dawson}, {Eftekharzadeh}, {Fernandez-Trincado}, {Gao},
  {Hutchinson}, {Jia}, {Jiang}, {Oravetz}, {Pan}, {Paris}, {Ponder}, {Peters},
  {Rogerson}, {Simmons}, {Smith}, \& {Wang}}]{Grier_etal_2017}
{Grier}, C.~J., {Trump}, J.~R., {Shen}, Y., {et~al.} 2017, \apj, 851, 21,
  \dodoi{10.3847/1538-4357/aa98dc}

\bibitem[{{Grier} {et~al.}(2019){Grier}, {Shen}, {Horne}, {Brandt}, {Trump},
  {Hall}, {Kinemuchi}, {Starkey}, {Schneider}, {Ho}, {Homayouni}, {I-Hsiu Li},
  {McGreer}, {Peterson}, {Bizyaev}, {Chen}, {Dawson}, {Eftekharzadeh}, {Guo},
  {Jia}, {Jiang}, {Kneib}, {Li}, {Li}, {Nie}, {Oravetz}, {Oravetz}, {Pan},
  {Petitjean}, {Ponder}, {Rogerson}, {Vivek}, {Zhang}, \&
  {Zou}}]{Grier_etal_2019}
{Grier}, C.~J., {Shen}, Y., {Horne}, K., {et~al.} 2019, \apj, 887, 38,
  \dodoi{10.3847/1538-4357/ab4ea5}

\bibitem[{{Hern{\'a}ndez Santisteban} {et~al.}(2020){Hern{\'a}ndez
  Santisteban}, {Edelson}, {Horne}, {Gelbord}, {Barth}, {Cackett}, {Goad},
  {Netzer}, {Starkey}, {Uttley}, {Brandt}, {Korista}, {Lohfink}, {Onken},
  {Page}, {Siegel}, {Vestergaard}, {Bisogni}, {Breeveld}, {Cenko}, {Dalla
  Bont{\`a}}, {Evans}, {Ferland}, {Gonzalez-Buitrago}, {Grupe}, {Joner},
  {Kriss}, {LaPorte}, {Mathur}, {Marshall}, {Mehdipour}, {Mudd}, {Peterson},
  {Schmidt}, {Vaughan}, \& {Valenti}}]{HernandezSantisteban_etal_2020}
{Hern{\'a}ndez Santisteban}, J.~V., {Edelson}, R., {Horne}, K., {et~al.} 2020,
  \mnras, 498, 5399, \dodoi{10.1093/mnras/staa2365}

\bibitem[{Hochreiter \& Schmidhuber(1997)}]{LSTM}
Hochreiter, S., \& Schmidhuber, J. 1997, Neural computation, 9, 1735

\bibitem[{{Homayouni} {et~al.}(2019){Homayouni}, {Trump}, {Grier}, {Shen},
  {Starkey}, {Brandt}, {Fonseca Alvarez}, {Hall}, {Horne}, {Kinemuchi}, {I-Hsiu
  Li}, {McGreer}, {Sun}, {Ho}, \& {Schneider}}]{Homayouni_etal_2019}
{Homayouni}, Y., {Trump}, J.~R., {Grier}, C.~J., {et~al.} 2019, \apj, 880, 126,
  \dodoi{10.3847/1538-4357/ab2638}

\bibitem[{{Homayouni} {et~al.}(2020){Homayouni}, {Trump}, {Grier}, {Horne},
  {Shen}, {Brandt}, {Dawson}, {Alvarez}, {Green}, {Hall}, {Hern{\'a}ndez
  Santisteban}, {Ho}, {Kinemuchi}, {Kochanek}, {Li}, {Peterson}, {Schneider},
  {Starkey}, {Bizyaev}, {Pan}, {Oravetz}, \& {Simmons}}]{Homayouni_etal_2020}
---. 2020, \apj, 901, 55, \dodoi{10.3847/1538-4357/ababa9}

\bibitem[{{Ivezi{\'c}} {et~al.}(2019){Ivezi{\'c}}, {Kahn}, {Tyson}, {Abel},
  {Acosta}, {Allsman}, {Alonso}, {AlSayyad}, {Anderson}, {Andrew}, {Angel},
  {Angeli}, {Ansari}, {Antilogus}, {Araujo}, {Armstrong}, {Arndt}, {Astier},
  {Aubourg}, {Auza}, {Axelrod}, {Bard}, {Barr}, {Barrau}, {Bartlett}, {Bauer},
  {Bauman}, {Baumont}, {Bechtol}, {Bechtol}, {Becker}, {Becla}, {Beldica},
  {Bellavia}, {Bianco}, {Biswas}, {Blanc}, {Blazek}, {Blandford}, {Bloom},
  {Bogart}, {Bond}, {Booth}, {Borgland}, {Borne}, {Bosch}, {Boutigny},
  {Brackett}, {Bradshaw}, {Brandt}, {Brown}, {Bullock}, {Burchat}, {Burke},
  {Cagnoli}, {Calabrese}, {Callahan}, {Callen}, {Carlin}, {Carlson},
  {Chandrasekharan}, {Charles-Emerson}, {Chesley}, {Cheu}, {Chiang}, {Chiang},
  {Chirino}, {Chow}, {Ciardi}, {Claver}, {Cohen-Tanugi}, {Cockrum}, {Coles},
  {Connolly}, {Cook}, {Cooray}, {Covey}, {Cribbs}, {Cui}, {Cutri}, {Daly},
  {Daniel}, {Daruich}, {Daubard}, {Daues}, {Dawson}, {Delgado}, {Dellapenna},
  {de Peyster}, {de Val-Borro}, {Digel}, {Doherty}, {Dubois},
  {Dubois-Felsmann}, {Durech}, {Economou}, {Eifler}, {Eracleous}, {Emmons},
  {Fausti Neto}, {Ferguson}, {Figueroa}, {Fisher-Levine}, {Focke}, {Foss},
  {Frank}, {Freemon}, {Gangler}, {Gawiser}, {Geary}, {Gee}, {Geha}, {Gessner},
  {Gibson}, {Gilmore}, {Glanzman}, {Glick}, {Goldina}, {Goldstein}, {Goodenow},
  {Graham}, {Gressler}, {Gris}, {Guy}, {Guyonnet}, {Haller}, {Harris},
  {Hascall}, {Haupt}, {Hernandez}, {Herrmann}, {Hileman}, {Hoblitt}, {Hodgson},
  {Hogan}, {Howard}, {Huang}, {Huffer}, {Ingraham}, {Innes}, {Jacoby}, {Jain},
  {Jammes}, {Jee}, {Jenness}, {Jernigan}, {Jevremovi{\'c}}, {Johns}, {Johnson},
  {Johnson}, {Jones}, {Juramy-Gilles}, {Juri{\'c}}, {Kalirai}, {Kallivayalil},
  {Kalmbach}, {Kantor}, {Karst}, {Kasliwal}, {Kelly}, {Kessler}, {Kinnison},
  {Kirkby}, {Knox}, {Kotov}, {Krabbendam}, {Krughoff}, {Kub{\'a}nek},
  {Kuczewski}, {Kulkarni}, {Ku}, {Kurita}, {Lage}, {Lambert}, {Lange},
  {Langton}, {Le Guillou}, {Levine}, {Liang}, {Lim}, {Lintott}, {Long},
  {Lopez}, {Lotz}, {Lupton}, {Lust}, {MacArthur}, {Mahabal}, {Mandelbaum},
  {Markiewicz}, {Marsh}, {Marshall}, {Marshall}, {May}, {McKercher}, {McQueen},
  {Meyers}, {Migliore}, {Miller}, {Mills}, {Miraval}, {Moeyens}, {Moolekamp},
  {Monet}, {Moniez}, {Monkewitz}, {Montgomery}, {Morrison}, {Mueller},
  {Muller}, {Mu{\~n}oz Arancibia}, {Neill}, {Newbry}, {Nief}, {Nomerotski},
  {Nordby}, {O'Connor}, {Oliver}, {Olivier}, {Olsen}, {O'Mullane}, {Ortiz},
  {Osier}, {Owen}, {Pain}, {Palecek}, {Parejko}, {Parsons}, {Pease},
  {Peterson}, {Peterson}, {Petravick}, {Libby Petrick}, {Petry},
  {Pierfederici}, {Pietrowicz}, {Pike}, {Pinto}, {Plante}, {Plate}, {Plutchak},
  {Price}, {Prouza}, {Radeka}, {Rajagopal}, {Rasmussen}, {Regnault}, {Reil},
  {Reiss}, {Reuter}, {Ridgway}, {Riot}, {Ritz}, {Robinson}, {Roby}, {Roodman},
  {Rosing}, {Roucelle}, {Rumore}, {Russo}, {Saha}, {Sassolas}, {Schalk},
  {Schellart}, {Schindler}, {Schmidt}, {Schneider}, {Schneider}, {Schoening},
  {Schumacher}, {Schwamb}, {Sebag}, {Selvy}, {Sembroski}, {Seppala}, {Serio},
  {Serrano}, {Shaw}, {Shipsey}, {Sick}, {Silvestri}, {Slater}, {Smith},
  {Smith}, {Sobhani}, {Soldahl}, {Storrie-Lombardi}, {Stover}, {Strauss},
  {Street}, {Stubbs}, {Sullivan}, {Sweeney}, {Swinbank}, {Szalay}, {Takacs},
  {Tether}, {Thaler}, {Thayer}, {Thomas}, {Thornton}, {Thukral}, {Tice},
  {Trilling}, {Turri}, {Van Berg}, {Vanden Berk}, {Vetter}, {Virieux},
  {Vucina}, {Wahl}, {Walkowicz}, {Walsh}, {Walter}, {Wang}, {Wang}, {Warner},
  {Wiecha}, {Willman}, {Winters}, {Wittman}, {Wolff}, {Wood-Vasey}, {Wu},
  {Xin}, {Yoachim}, \& {Zhan}}]{Ivezic_etal_2019}
{Ivezi{\'c}}, {\v{Z}}., {Kahn}, S.~M., {Tyson}, J.~A., {et~al.} 2019, \apj,
  873, 111, \dodoi{10.3847/1538-4357/ab042c}

\bibitem[{{Jiang} {et~al.}(2017){Jiang}, {Green}, {Greene}, {Morganson},
  {Shen}, {Pancoast}, {MacLeod}, {Anderson}, {Brandt}, {Grier}, {Rix}, {Ruan},
  {Protopapas}, {Scott}, {Burgett}, {Hodapp}, {Huber}, {Kaiser}, {Kudritzki},
  {Magnier}, {Metcalfe}, {Tonry}, {Wainscoat}, \& {Waters}}]{Jiang_etal_2017}
{Jiang}, Y.-F., {Green}, P.~J., {Greene}, J.~E., {et~al.} 2017, \apj, 836, 186,
  \dodoi{10.3847/1538-4357/aa5b91}

\bibitem[{{Kasliwal} {et~al.}(2015){Kasliwal}, {Vogeley}, \&
  {Richards}}]{Kasliwal_etal_2015}
{Kasliwal}, V.~P., {Vogeley}, M.~S., \& {Richards}, G.~T. 2015, \mnras, 451,
  4328, \dodoi{10.1093/mnras/stv1230}

\bibitem[{{Kelly} {et~al.}(2009){Kelly}, {Bechtold}, \&
  {Siemiginowska}}]{Kelly_etal_2009}
{Kelly}, B.~C., {Bechtold}, J., \& {Siemiginowska}, A. 2009, \apj, 698, 895,
  \dodoi{10.1088/0004-637X/698/1/895}

\bibitem[{{Kelly} {et~al.}(2011){Kelly}, {Sobolewska}, \&
  {Siemiginowska}}]{Kelly_etal_2011}
{Kelly}, B.~C., {Sobolewska}, M., \& {Siemiginowska}, A. 2011, \apj, 730, 52,
  \dodoi{10.1088/0004-637X/730/1/52}

\bibitem[{{Koz{\l}owski} {et~al.}(2010){Koz{\l}owski}, {Kochanek}, {Udalski},
  {Wyrzykowski}, {Soszy{\'n}ski}, {Szyma{\'n}ski}, {Kubiak}, {Pietrzy{\'n}ski},
  {Szewczyk}, {Ulaczyk}, {Poleski}, \& {OGLE
  Collaboration}}]{Kozlowski_etal_2010}
{Koz{\l}owski}, S., {Kochanek}, C.~S., {Udalski}, A., {et~al.} 2010, \apj, 708,
  927, \dodoi{10.1088/0004-637X/708/2/927}

\bibitem[{{Li} {et~al.}(2019){Li}, {Shen}, {Brandt}, {Grier}, {Hall}, {Ho},
  {Homayouni}, {Horne}, {Schneider}, {Trump}, \& {Starkey}}]{Li_etal_2019}
{Li}, I-Hsiu, J., {Shen}, Y., {Brandt}, W.~N., {et~al.} 2019, \apj, 884, 119,
  \dodoi{10.3847/1538-4357/ab41fb}

\bibitem[{{Liao} {et~al.}(2021){Liao}, {Chen}, {Liu}, {Holgado}, {Guo},
  {Gruendl}, {Morganson}, {Shen}, {Davis}, {Kessler}, {Martini}, {McMahon},
  {Allam}, {Annis}, {Avila}, {Banerji}, {Bechtol}, {Bertin}, {Brooks},
  {Buckley-Geer}, {Carnero Rosell}, {Carrasco Kind}, {Carretero}, {Javier
  Castander}, {Cunha}, {D'Andrea}, {da Costa}, {Davis}, {De Vicente}, {Desai},
  {Thomas Diehl}, {Doel}, {Eifler}, {Evrard}, {Flaugher}, {Fosalba}, {Frieman},
  {Garcia-Bellido}, {Gaztanaga}, {Glazebrook}, {Gruen}, {Gschwend},
  {Gutierrez}, {Hartley}, {Hollowood}, {Honscheid}, {Hoyle}, {James}, {Krause},
  {Kuehn}, {Lima}, {Maia}, {Marshall}, {Menanteau}, {Miquel}, {Plazas
  Malag{\'o}n}, {Roodman}, {Sanchez}, {Scarpine}, {Schubnell}, {Serrano},
  {Smith}, {Smith}, {Soares-Santos}, {Sobreira}, {Suchyta}, {Swanson}, {Tarle},
  {Vikram}, \& {Walker}}]{Liao_etal_2021}
{Liao}, W.-T., {Chen}, Y.-C., {Liu}, X., {et~al.} 2021, \mnras, 500, 4025,
  \dodoi{10.1093/mnras/staa3055}

\bibitem[{{MacLeod} {et~al.}(2010){MacLeod}, {Ivezi{\'c}}, {Kochanek},
  {Koz{\l}owski}, {Kelly}, {Bullock}, {Kimball}, {Sesar}, {Westman}, {Brooks},
  {Gibson}, {Becker}, \& {de Vries}}]{MacLeod_etal_2010}
{MacLeod}, C.~L., {Ivezi{\'c}}, {\v{Z}}., {Kochanek}, C.~S., {et~al.} 2010,
  \apj, 721, 1014, \dodoi{10.1088/0004-637X/721/2/1014}

\bibitem[{{MacLeod} {et~al.}(2012){MacLeod}, {Ivezi{\'c}}, {Sesar}, {de Vries},
  {Kochanek}, {Kelly}, {Becker}, {Lupton}, {Hall}, {Richards}, {Anderson}, \&
  {Schneider}}]{MacLeod_etal_2012}
{MacLeod}, C.~L., {Ivezi{\'c}}, {\v{Z}}., {Sesar}, B., {et~al.} 2012, \apj,
  753, 106, \dodoi{10.1088/0004-637X/753/2/106}

\bibitem[{{McHardy} {et~al.}(2018){McHardy}, {Connolly}, {Horne}, {Cackett},
  {Gelbord}, {Peterson}, {Pahari}, {Gehrels}, {Goad}, {Lira}, {Arevalo},
  {Baldi}, {Brandt}, {Breedt}, {Chand}, {Dewangan}, {Done}, {Elvis},
  {Emmanoulopoulos}, {Fausnaugh}, {Kaspi}, {Kochanek}, {Korista}, {Papadakis},
  {Rao}, {Uttley}, {Vestergaard}, \& {Ward}}]{McHardy_etal_2018}
{McHardy}, I.~M., {Connolly}, S.~D., {Horne}, K., {et~al.} 2018, \mnras, 480,
  2881, \dodoi{10.1093/mnras/sty1983}

\bibitem[{{Moreno} {et~al.}(2021){Moreno}, {Buttry}, {O'Brien}, {Vogeley},
  {Richards}, \& {Smith}}]{Moreno_etal_2021}
{Moreno}, J., {Buttry}, R., {O'Brien}, J., {et~al.} 2021, \aj, 162, 232,
  \dodoi{10.3847/1538-3881/ac205c}

\bibitem[{{Mudd} {et~al.}(2018){Mudd}, {Martini}, {Zu}, {Kochanek}, {Peterson},
  {Kessler}, {Davis}, {Hoormann}, {King}, {Lidman}, {Sommer}, {Tucker},
  {Asorey}, {Hinton}, {Glazebrook}, {Kuehn}, {Lewis}, {Macaulay}, {Moeller},
  {O'Neill}, {Zhang}, {Abbott}, {Abdalla}, {Allam}, {Banerji},
  {Benoit-L{\'e}vy}, {Bertin}, {Brooks}, {Carnero Rosell}, {Carollo}, {Carrasco
  Kind}, {Carretero}, {Cunha}, {D'Andrea}, {da Costa}, {Davis}, {Desai},
  {Doel}, {Fosalba}, {Garc{\'\i}a-Bellido}, {Gaztanaga}, {Gerdes}, {Gruen},
  {Gruendl}, {Gschwend}, {Gutierrez}, {Hartley}, {Honscheid}, {James},
  {Kuhlmann}, {Kuropatkin}, {Lima}, {Maia}, {Marshall}, {McMahon}, {Menanteau},
  {Miquel}, {Plazas}, {Romer}, {Sanchez}, {Schindler}, {Schubnell}, {Smith},
  {Smith}, {Soares-Santos}, {Sobreira}, {Suchyta}, {Swanson}, {Tarle},
  {Thomas}, {Tucker}, {Walker}, \& {DES Collaboration}}]{Mudd_etal_2018}
{Mudd}, D., {Martini}, P., {Zu}, Y., {et~al.} 2018, \apj, 862, 123,
  \dodoi{10.3847/1538-4357/aac9bb}

\bibitem[{{Neustadt} \& {Kochanek}(2022)}]{Neustadt_etal_2022}
{Neustadt}, J.~M.~M., \& {Kochanek}, C.~S. 2022, \mnras, 513, 1046,
  \dodoi{10.1093/mnras/stac888}

\bibitem[{{Neustadt} {et~al.}(2024){Neustadt}, {Kochanek}, {Montano},
  {Gelbord}, {Barth}, {De Rosa}, {Kriss}, {Cackett}, {Horne}, {Kara}, {Landt},
  {Netzer}, {Arav}, {Bentz}, {Dalla Bont{\`a}}, {Dehghanian}, {Du}, {Edelson},
  {Ferland}, {Fian}, {Fischer}, {Goad}, {Gonz{\'a}lez Buitrago}, {Gorjian},
  {Grier}, {Hall}, {Homayouni}, {Hu}, {Ili{\'c}}, {Joner}, {Kaastra}, {Kaspi},
  {Korista}, {Kova{\v{c}}evi{\'c}}, {Lewin}, {Li}, {McHardy}, {Mehdipour},
  {Miller}, {Panagiotou}, {Partington}, {Plesha}, {Pogge}, {Popovi{\'c}},
  {Proga}, {Storchi-Bergmann}, {Sanmartim}, {Siebert}, {Signorini},
  {Vestergaard}, {Zaidouni}, \& {Zu}}]{Neustadt_etal_2024}
{Neustadt}, J. M.~M., {Kochanek}, C.~S., {Montano}, J., {et~al.} 2024, \apj,
  961, 219, \dodoi{10.3847/1538-4357/ad1386}

\bibitem[{{Ni} {et~al.}(2021){Ni}, {Brandt}, {Chen}, {Luo}, {Nyland}, {Yang},
  {Zou}, {Aird}, {Alexander}, {Bauer}, {Lacy}, {Lehmer}, {Mallick}, {Salvato},
  {Schneider}, {Tozzi}, {Traulsen}, {Vaccari}, {Vignali}, {Vito}, {Xue},
  {Banerji}, {Chow}, {Comastri}, {Del Moro}, {Gilli}, {Mullaney}, {Paolillo},
  {Schwope}, {Shemmer}, {Sun}, {Timlin}, \& {Trump}}]{Ni_etal_2021a}
{Ni}, Q., {Brandt}, W.~N., {Chen}, C.-T., {et~al.} 2021, \apjs, 256, 21,
  \dodoi{10.3847/1538-4365/ac0dc6}

\bibitem[{Papamakarios {et~al.}(2017)Papamakarios, Pavlakou, \&
  Murray}]{Papamakarios_etal_2017}
Papamakarios, G., Pavlakou, T., \& Murray, I. 2017, Advances in neural
  information processing systems, 30

\bibitem[{{Peterson} {et~al.}(1998){Peterson}, {Wanders}, {Horne}, {Collier},
  {Alexander}, {Kaspi}, \& {Maoz}}]{Peterson_etal_1998}
{Peterson}, B.~M., {Wanders}, I., {Horne}, K., {et~al.} 1998, \pasp, 110, 660,
  \dodoi{10.1086/316177}

\bibitem[{{S{\'a}nchez-S{\'a}ez} {et~al.}(2021){S{\'a}nchez-S{\'a}ez}, {Lira},
  {Mart{\'\i}}, {S{\'a}nchez-Pi}, {Arredondo}, {Bauer}, {Bayo},
  {Cabrera-Vives}, {Donoso-Oliva}, {Est{\'e}vez}, {Eyheramendy}, {F{\"o}rster},
  {Hern{\'a}ndez-Garc{\'\i}a}, {Arancibia}, {P{\'e}rez-Carrasco},
  {Sep{\'u}lveda}, \& {Vergara}}]{Sanchez-saez_etal_2021}
{S{\'a}nchez-S{\'a}ez}, P., {Lira}, H., {Mart{\'\i}}, L., {et~al.} 2021, \aj,
  162, 206, \dodoi{10.3847/1538-3881/ac1426}

\bibitem[{{Secunda} {et~al.}(2023){Secunda}, {Jiang}, \&
  {Greene}}]{Secunda_etal_2023}
{Secunda}, A., {Jiang}, Y.-F., \& {Greene}, J.~E. 2023, arXiv e-prints,
  arXiv:2311.10820, \dodoi{10.48550/arXiv.2311.10820}

\bibitem[{{Sergeev} {et~al.}(2005){Sergeev}, {Doroshenko}, {Golubinskiy},
  {Merkulova}, \& {Sergeeva}}]{Sergeev_etal_2005}
{Sergeev}, S.~G., {Doroshenko}, V.~T., {Golubinskiy}, Y.~V., {Merkulova},
  N.~I., \& {Sergeeva}, E.~A. 2005, \apj, 622, 129, \dodoi{10.1086/427820}

\bibitem[{{Shakura} \& {Sunyaev}(1973)}]{Shakura_Sunyave_1973}
{Shakura}, N.~I., \& {Sunyaev}, R.~A. 1973, \aap, 24, 337

\bibitem[{{Starkey} {et~al.}(2017){Starkey}, {Horne}, {Fausnaugh}, {Peterson},
  {Bentz}, {Kochanek}, {Denney}, {Edelson}, {Goad}, {De Rosa}, {Anderson},
  {Ar{\'e}valo}, {Barth}, {Bazhaw}, {Borman}, {Boroson}, {Bottorff}, {Brandt},
  {Breeveld}, {Cackett}, {Carini}, {Croxall}, {Crenshaw}, {Dalla Bont{\`a}},
  {De Lorenzo-C{\'a}ceres}, {Dietrich}, {Efimova}, {Ely}, {Evans},
  {Filippenko}, {Flatland}, {Gehrels}, {Geier}, {Gelbord}, {Gonzalez},
  {Gorjian}, {Grier}, {Grupe}, {Hall}, {Hicks}, {Horenstein}, {Hutchison},
  {Im}, {Jensen}, {Joner}, {Jones}, {Kaastra}, {Kaspi}, {Kelly}, {Kennea},
  {Kim}, {Kim}, {Klimanov}, {Korista}, {Kriss}, {Lee}, {Leonard}, {Lira},
  {MacInnis}, {Manne-Nicholas}, {Mathur}, {McHardy}, {Montouri}, {Musso},
  {Nazarov}, {Norris}, {Nousek}, {Okhmat}, {Pancoast}, {Parks}, {Pei}, {Pogge},
  {Pott}, {Rafter}, {Rix}, {Saylor}, {Schimoia}, {Schn{\"u}lle}, {Sergeev},
  {Siegel}, {Spencer}, {Sung}, {Teems}, {Turner}, {Uttley}, {Vestergaard},
  {Villforth}, {Weiss}, {Woo}, {Yan}, {Young}, {Zheng}, \&
  {Zu}}]{Starkey_etal_2017}
{Starkey}, D., {Horne}, K., {Fausnaugh}, M.~M., {et~al.} 2017, \apj, 835, 65,
  \dodoi{10.3847/1538-4357/835/1/65}

\bibitem[{{Starkey} {et~al.}(2016){Starkey}, {Horne}, \&
  {Villforth}}]{Starkey_etal_2016}
{Starkey}, D.~A., {Horne}, K., \& {Villforth}, C. 2016, \mnras, 456, 1960,
  \dodoi{10.1093/mnras/stv2744}

\bibitem[{{Starkey} {et~al.}(2023){Starkey}, {Huang}, {Horne}, \&
  {Lin}}]{Starkey_etal_2023}
{Starkey}, D.~A., {Huang}, J., {Horne}, K., \& {Lin}, D. N.~C. 2023, \mnras,
  519, 2754, \dodoi{10.1093/mnras/stac3579}

\bibitem[{{Stone} \& {Shen}(2023)}]{Stone_etal_2023}
{Stone}, Z., \& {Shen}, Y. 2023, \mnras, 524, 4521,
  \dodoi{10.1093/mnras/stad2034}

\bibitem[{{Stone} {et~al.}(2022){Stone}, {Shen}, {Burke}, {Chen}, {Yang},
  {Liu}, {Gruendl}, {Adam{\'o}w}, {Andrade-Oliveira}, {Annis}, {Bacon},
  {Bertin}, {Bocquet}, {Brooks}, {Burke}, {Carnero Rosell}, {Carrasco Kind},
  {Carretero}, {da Costa}, {Pereira}, {De Vicente}, {Desai}, {Diehl}, {Doel},
  {Ferrero}, {Friedel}, {Frieman}, {Garc{\'\i}a-Bellido}, {Gaztanaga}, {Gruen},
  {Gutierrez}, {Hinton}, {Hollowood}, {Honscheid}, {James}, {Kuehn},
  {Kuropatkin}, {Lidman}, {Maia}, {Menanteau}, {Miquel}, {Morgan},
  {Paz-Chinch{\'o}n}, {Pieres}, {Plazas Malag{\'o}n}, {Rodriguez-Monroy},
  {Sanchez}, {Scarpine}, {Serrano}, {Sevilla-Noarbe}, {Smith}, {Suchyta},
  {Swanson}, {Tarl{\'e}}, {To}, \& {DES Collaboration}}]{Stone_etal_2022}
{Stone}, Z., {Shen}, Y., {Burke}, C.~J., {et~al.} 2022, \mnras, 514, 164,
  \dodoi{10.1093/mnras/stac1259}

\bibitem[{{Sun} {et~al.}(2018){Sun}, {Grier}, \& {Peterson}}]{PyCCF}
{Sun}, M., {Grier}, C.~J., \& {Peterson}, B.~M. 2018, {PyCCF: Python Cross
  Correlation Function for reverberation mapping studies}, Astrophysics Source
  Code Library, record ascl:1805.032

\bibitem[{{Tachibana} {et~al.}(2020){Tachibana}, {Graham}, {Kawai},
  {Djorgovski}, {Drake}, {Mahabal}, \& {Stern}}]{Tachibana_etal_2020}
{Tachibana}, Y., {Graham}, M.~J., {Kawai}, N., {et~al.} 2020, \apj, 903, 54,
  \dodoi{10.3847/1538-4357/abb9a9}

\bibitem[{Talts {et~al.}(2018)Talts, Betancourt, Simpson, Vehtari, \&
  Gelman}]{Talts_etal_2018}
Talts, S., Betancourt, M., Simpson, D., Vehtari, A., \& Gelman, A. 2018, arXiv
  preprint arXiv:1804.06788

\bibitem[{Tejero-Cantero {et~al.}(2020)Tejero-Cantero, Boelts, Deistler,
  Lueckmann, Durkan, Gonçalves, Greenberg, \& Macke}]{sbi}
Tejero-Cantero, A., Boelts, J., Deistler, M., {et~al.} 2020, Journal of Open
  Source Software, 5, 2505, \dodoi{10.21105/joss.02505}

\bibitem[{{Vanden Berk} {et~al.}(2004){Vanden Berk}, {Wilhite}, {Kron},
  {Anderson}, {Brunner}, {Hall}, {Ivezi{\'c}}, {Richards}, {Schneider}, {York},
  {Brinkmann}, {Lamb}, {Nichol}, \& {Schlegel}}]{VandenBerk_etal_2004}
{Vanden Berk}, D.~E., {Wilhite}, B.~C., {Kron}, R.~G., {et~al.} 2004, \apj,
  601, 692, \dodoi{10.1086/380563}

\bibitem[{{Wang} {et~al.}(2023){Wang}, {Leja}, {Villar}, \&
  {Speagle}}]{Wang_etal_2023}
{Wang}, B., {Leja}, J., {Villar}, V.~A., \& {Speagle}, J.~S. 2023, \apjl, 952,
  L10, \dodoi{10.3847/2041-8213/ace361}

\bibitem[{{Yu} {et~al.}(2020){Yu}, {Martini}, {Davis}, {Gruendl}, {Hoormann},
  {Kochanek}, {Lidman}, {Mudd}, {Peterson}, {Wester}, {Allam}, {Annis},
  {Asorey}, {Avila}, {Banerji}, {Bertin}, {Brooks}, {Buckley-Geer}, {Calcino},
  {Rosell}, {Carollo}, {Kind}, {Carretero}, {Cunha}, {D'Andrea}, {Costa}, {De
  Vicente}, {Desai}, {Diehl}, {Doel}, {Eifler}, {Flaugher}, {Fosalba},
  {Frieman}, {Garc{\'\i}a-Bellido}, {Gaztanaga}, {Glazebrook}, {Gruen},
  {Gschwend}, {Gutierrez}, {Hartley}, {Hinton}, {Hollowood}, {Honscheid},
  {Hoyle}, {James}, {Kim}, {Krause}, {Kuehn}, {Kuropatkin}, {Lewis}, {Lima},
  {Macaulay}, {Maia}, {Marshall}, {Menanteau}, {Miquel}, {M{\"o}ller},
  {Plazas}, {Romer}, {Sanchez}, {Scarpine}, {Schubnell}, {Serrano}, {Smith},
  {Smith}, {Soares-Santos}, {Sobreira}, {Suchyta}, {Swann}, {Swanson}, {Tarle},
  {Tucker}, {Tucker}, \& {Vikram}}]{Yu_etal_2020}
{Yu}, Z., {Martini}, P., {Davis}, T.~M., {et~al.} 2020, \apjs, 246, 16,
  \dodoi{10.3847/1538-4365/ab5e7a}

\bibitem[{{Zaidouni} {et~al.}(2024){Zaidouni}, {Kara}, {Kosec}, {Mehdipour},
  {Rogantini}, {Kriss}, {Behar}, {Kaastra}, {Barth}, {Cackett}, {De Rosa},
  {Homayouni}, {Horne}, {Landt}, {Arav}, {Bentz}, {Brotherton}, {Dalla
  Bont{\`a}}, {Dehghanian}, {Ferland}, {Fian}, {Gelbord}, {Goad}, {Gonz{\'a}lez
  Buitrago}, {Grier}, {Hall}, {Hu}, {Ili{\'c}}, {Kaspi}, {Kochanek},
  {Kova{\v{c}}evi{\'c}}, {Kynoch}, {Lewin}, {Montano}, {Netzer}, {Neustadt},
  {Panagiotou}, {Partington}, {Plesha}, {Popovi{\'c}}, {Proga},
  {Storchi-Bergmann}, {Sanmartim}, {Siebert}, {Signorini}, {Vestergaard},
  {Waters}, \& {Zu}}]{Zaidouni_etal_2024}
{Zaidouni}, F., {Kara}, E., {Kosec}, P., {et~al.} 2024, arXiv e-prints,
  arXiv:2406.17061, \dodoi{10.48550/arXiv.2406.17061}

\bibitem[{{Zou} {et~al.}(2022){Zou}, {Brandt}, {Chen}, {Leja}, {Ni}, {Yan},
  {Yang}, {Zhu}, {Luo}, {Nyland}, {Vito}, \& {Xue}}]{Zou_etal_2022}
{Zou}, F., {Brandt}, W.~N., {Chen}, C.-T., {et~al.} 2022, \apjs, 262, 15,
  \dodoi{10.3847/1538-4365/ac7bdf}

\bibitem[{{Zu} {et~al.}(2016){Zu}, {Kochanek}, {Koz{\l}owski}, \&
  {Peterson}}]{Zu_etal_2016}
{Zu}, Y., {Kochanek}, C.~S., {Koz{\l}owski}, S., \& {Peterson}, B.~M. 2016,
  \apj, 819, 122, \dodoi{10.3847/0004-637X/819/2/122}

\bibitem[{{Zu} {et~al.}(2013){Zu}, {Kochanek}, {Koz{\l}owski}, \&
  {Udalski}}]{Zu_etal_2013}
{Zu}, Y., {Kochanek}, C.~S., {Koz{\l}owski}, S., \& {Udalski}, A. 2013, \apj,
  765, 106, \dodoi{10.1088/0004-637X/765/2/106}

\bibitem[{{Zu} {et~al.}(2011){Zu}, {Kochanek}, \& {Peterson}}]{Zu_etal_2011}
{Zu}, Y., {Kochanek}, C.~S., \& {Peterson}, B.~M. 2011, \apj, 735, 80,
  \dodoi{10.1088/0004-637X/735/2/80}

\end{thebibliography}

\end{document}